\documentclass[lettersize,journal]{IEEEtran}
\usepackage{amsmath,amsfonts}
\usepackage{algorithmic}
\usepackage{algorithm}
\usepackage{array}
\usepackage[caption=false,font=footnotesize,labelfont=rm,textfont=rm]{subfig}
\usepackage{textcomp}
\usepackage{stfloats}
\usepackage{url}
\usepackage{verbatim}
\usepackage{graphicx}
\usepackage{cite}
\usepackage{graphicx}
\usepackage{epstopdf}
\usepackage{subfloat}
\usepackage{enumerate}
\usepackage{float}
\usepackage{subfig}
\usepackage{multirow}
\usepackage{color}
\usepackage{hyperref}
\usepackage{booktabs}
\usepackage{graphicx}
\usepackage[justification=centering]{caption}
\usepackage[font=scriptsize]{caption}
\usepackage{tikz}
\usepackage{bigstrut,multirow,rotating,booktabs}
\usepackage{pifont}
\hypersetup{
	colorlinks=true,
	linkcolor=red,
	filecolor=blue,      
	urlcolor=blue,
	citecolor=blue,
}
\newcommand*{\circled}[1]{\lower.7ex\hbox{\tikz\draw (0pt, 0pt)%
    circle (.5em) node {\makebox[1em][c]{\small #1}};}}

\begin{document}
\title{DT-JRD: Deep Transformer based Just Recognizable Difference Prediction Model for Video Coding for Machines}
\author{Junqi Liu, Yun Zhang,~\IEEEmembership{Senior Member,~IEEE}, Xiaoqi Wang, Xu Long, Sam Kwong, \IEEEmembership{Fellow, IEEE}
\thanks{Junqi Liu, Yun Zhang, and Xiaoqi Wang are with the School of Electronics and Communication Engineering, Shenzhen Campus, Sun Yat-Sen University, Shenzhen 518017, China. (Email: liujq85@mail2.sysu.edu.cn, {zhangyun2}@mail.sysu.edu.cn, wangxq79@mail2.sysu.edu.cn).}
\thanks{X. Long is with the National Astronomical Observatories, Chinese Academy of Sciences, Beijing, 100012, China. (Email:lxu@nao.cas.cn).}
\thanks{S. Kwong is with the Department of Computing and Decision Sciences, Lingnan University, Hong Kong (e-mail: samkwong@ln.edu.hk).}

}
\markboth{IEEE Transactions on Multimedia}%
{
Lin \MakeLowercase{\textit{\textit{et al.}}}:
}


\maketitle

\begin{abstract}
Just Recognizable Difference (JRD) represents the minimum visual difference that is detectable by machine vision, which can be exploited to promote machine vision oriented visual signal processing. In this paper, we propose a Deep Transformer based JRD (DT-JRD) prediction model for Video Coding for Machines (VCM), where the accurately predicted JRD can be used reduce the coding bit rate while maintaining the accuracy of machine tasks.
Firstly, we model the JRD prediction as a multi-class classification and propose a DT-JRD prediction model that integrates an improved embedding, a content and distortion feature extraction, a multi-class classification and a novel learning strategy. 
Secondly, inspired by the perception property that machine vision exhibits a similar response to distortions near JRD, we propose an asymptotic JRD loss by using Gaussian Distribution-based Soft Labels (GDSL), which significantly extends the number of training labels and relaxes classification boundaries.
Finally, we propose a DT-JRD based VCM to reduce the coding bits while maintaining the accuracy of object detection.
Extensive experimental results demonstrate that the mean absolute error of the predicted JRD by the DT-JRD is 5.574, outperforming the state-of-the-art JRD prediction model by 13.1\%.
Coding experiments shows that comparing with the VVC, the DT-JRD based VCM achieves an average of 29.58\% bit rate reduction while maintaining the object detection accuracy.
\end{abstract}


\section{Introduction}
\IEEEPARstart{W}{ith} the rapid development of Artificial Intelligence (AI) and multimedia communication, remote intelligent visual analytics \cite{ren2015faster} is highly demanded for smart city, privacy preserving, robotics and remote control, etc..
Video/Image Coding for Machine (VCM/ICM) \cite{duan2020video} that aims to encode visual content more efficiently for remote visual analysis is highly demanded for machine-to-machine visual communication. The main challenge in the VCM field is to reduce coding bit rate while maintaining high performance in machine vision tasks.

Conventionally, there are a number of visual coding standards, such as JPEG\cite{wallace1992jpeg} and JPEG2000\cite{skodras2001jpeg} for image compression and High Efficiency Video Coding (HEVC)\cite{sullivan2012overview} and Versatile Video Coding (VVC)\cite{bross2021overview} for video compression, which exploited spatial-temporal correlation, symbol redundancies as well as the visual redundancies in Human Visual System (HVS).
With the vigorous development of deep learning, a number of Learned Image Compression (LIC) methods have been developed, such as Non-Local Attention optimization and Improved Context modeling-based image compression (NLAIC)\cite{chen2021end}, Efficient Learned Image Compression (ELIC)\cite{he2022elic} and Multi-reference entropy model based Learned Image Compression (MLIC)\cite{jiang2023mlic}, which surpassed traditional image coding with deep network architecture and advanced entropy model. Thanks to the powerful feature extraction and representation abilities, the LIC represented the visual content well and exploited the redundancies effectively for human vision. Both traditional and learned image/video codecs mainly leveraged visual properties of HVS and aimed to minimize the image/video quality degradation at given bit rate constraints.

Due to the contrast and brightness visual sensitivities, and spatial-temporal and pattern masking effects in HVS, distortion below the Just Noticeable Difference (JND) are not perceivable. Based on the understandings of the visual properties in HVS, many JND estimation models have been proposed, which mainly include pixel-domain JND model\cite{jiang2022toward}, transform-domain JND model, block-level JND model\cite{tian2021perceptual,9810507}, and picture/video-wise JND model\cite{tian2020just,shen2020just,liu2019deep,zhang2021deep} .
Liu \textit{et al.}\cite{liu2019deep} modeled Picture-Wise JND (PW-JND) prediction as a multi-class classification which was solved by using multiple binary classifications. Moreover, a deep learning-based binary predictor was proposed to predict whether the distortion of an image was visually perceivable or not as compared to its reference image.
By additionally fusing the temporal information extracted from the optical flow of the video, Zhang \textit{et al.} \cite{zhang2021deep} proposed a Video Wise Spatial-Temporal Satisfied User Ratio (VW-STSUR) model where the JND of the video was determined as 75\% of the Satisfied User Ratio (SUR). 
Nami \textit{et al.}\cite{10500870} formulated JND prediction as a multi-task problem, where JND prediction was jointly learned with reconstructing JND-quality frames. Furthermore, three JND levels in HVS were learned simultaneously, which not only improved the prediction accuracy but also reduced the complexity.
As one of the most important visual mechanisms in HVS, JND inspires perceptually optimized coding methods \cite{zhang2023survey,zhang2019perceptual} which leverage the visual redundancies and improve the perceptual quality .
Zhou \textit{et al.}\cite{zhou2020just} proposed a Rate-Distortion (R-D) model based on the JND factor and developed a rate control approach for HEVC.
Zhu \textit{et al.}\cite{zhu2019jnd} incorporated the pixel-wise JND model into the Rate Distortion Optimization (RDO) for the AV1 encoder and proposed a perceptual distortion measure combining Mean Squared Error (MSE) and JND to adjust the Lagrange multiplier.
With breakthroughs in end-to-end image compression networks, JND was employed in their perceptual optimization as well.
Ding \textit{et al.}\cite{ding2023jnd} designed a JND-based perceptual quality loss and a distortion-aware adjuster, outperforming the baseline end-to-end networks.
Based on the JND dataset Videoset\cite{wang2017videoset}, Pakdaman \textit{et al.}\cite{pakdaman2024perceptual} proved that image-wise JND loss and feature-wise JND loss helped to achieve an improvement in rate-distortion performance for learned image compression.
These works were proposed to model or exploit the JND in HVS. However, they can hardly be applicable to VCM as they did not consider the characteristics of machine vision.

Recently, VCM aims to minimize the coding bit rate while maintaining the performance of downstream machine vision tasks. VCM becomes a hot research topic in academia and industry.
With the development of learning based image/video compression, a number of works on end-to-end VCM networks have been developed.
Le \textit{et al.}\cite{le2021image} proposed a machine-oriented learned image codec and a dynamic loss weighting strategy to balance rate loss, distortion loss, and machine vision task loss.
Gao \textit{et al.}\cite{gao2021towards} proposed semantic-oriented metrics for developing LIC in multiple machine vision tasks.
Choi \textit{et al.}\cite{choi2022scalable} designed a scalable hierarchical codec that used different levels of features in the latent space for multiple tasks, including reconstruction, detection, and segmentation.
Chen \textit{et al.}\cite{chen2023transtic} applied visual prompts on Transformer-based LIC, which achieved remarkable transferable performance on various machine tasks.
Feng \textit{et al.}\cite{feng2022image} proposed an omnipotent feature learning framework based on contrastive learning, and constrained entropy by an information filtering module to make the framework more suitable for feature compression.
These approaches optimized the VCM in an end-to-end way from network architectures, optimization objectives, multi-task perspective, and so on. 
However, the end-to-end implicit feature learning in these VCM approaches did not effectively exploit machine vision properties.

To further optimize the VCM, it is important to reveal and exploit the perceptual mechanism of machine vision.
Recent studies revealed that machine vision has different wider frequency bandwidth\cite{subramanian2024spatial}, higher contrast sensitivity \cite{chiu2022humanvisualcontrastsensitivity}, and lower robustness \cite{geirhos2021partial} as compared with human vision.
On contrary, machine vision also has attention mechanisms similar to the visual attention, visual masking, and non-local constraints in human vision. Moreover, it was found that there is a Just Recognizable Difference (JRD) for machine vision\cite{zhang2021just}, indicating the minimum distortion that affects the performance of machine vision significantly. To investigate the JRD,
Jin \textit{et al.} \cite{jin2021just} proposed a Deep Machine Vision-Just Noticeable Difference (DMV-JND) model to find visual redundant features for each pixel in an image.
With the assistance of the magnitude and spatial restraint strategies from DMV-JND, distorted images with a lower average Peak Signal-to-Noise Ratio (PSNR) were generated while maintaining the classification accuracy at an acceptable level.
Akan \textit{et al.}\cite{akan2022just} proposed a machine-perceived JND measurement and an adversarial image generation algorithm that added noise continuously to an image until the model output incorrect labels, deceiving the machine vision model.
Motivated by the concept of SUR, Zhang \textit{et al.}\cite{zhang2022smr} established a large-scale Satisfied Machine Ratio (SMR) dataset, which contained 72 and 98 models for classification and detection respectively. Based on the SMR dataset, they proposed to predict the SMR for perceptual VCM. This work took into account the properties of diverse machines and was conducive to general ICM/VCM.
Zhang \textit{et al.}\cite{zhang2021just} proposed an ensemble-learning-based JRD prediction framework for object detection. By decomposing multi-classification into 8 binary classifiers, coarse-grained prediction results were searched to get the predicted JRD.
Zhang \textit{et al.}\cite{zhang2023learning} built an Object-Wise JRD (OW-JRD) dataset with 29218 objects and each object image has 64 distortion levels from VVC intra coding. Then, they proposed a EfficientNet-based OW-JRD prediction model to predict the JRD from 64 classes, which was decomposed into multiple binary classifications by using binary classifiers. Nevertheless, this method required multiple times of image compression and prediction, which caused additional complexity overheads. Also, the JRD prediction accuracy could be further improved.

In this paper, we propose a Deep Transformer-based JRD (DT-JRD) prediction model for VCM. The contributions are
\begin{enumerate}
\item  We model the JRD prediction as a multi-class classification problem and propose a novel DT-JRD framework, which integrates an improved embedding, a content and distortion feature extraction, a multi-class classification, and a novel learning strategy.
\item To improve the DT-JRD model training, we propose an asymptotic JRD loss by using Gaussian Distribution-based Soft Labels (GDSL), which extends the number of training labels and relaxes classification boundaries.
\item We propose a DT-JRD based VCM optimization, which reduces the coding bits while maintaining the accuracy of object detection. 
\end{enumerate}

The remainder of this paper is organized as follows. Section \ref{section:motivation} describes the motivation and problem of JRD prediction. Section \ref{section:framework} to Section \ref{section:GDSL} presents the framework and key modules of DT-JRD. Section \ref{section:coding} introduces the DT-JRD based VCM method. Section \ref{section:experiment} presents the experimental results and analysis. Section \ref{section:conclusion} draws the conclusion.

\section{The DT-JRD Prediction Model for VCM}
\begin{figure}[t]
	\centering
        \includegraphics[width=0.5\textwidth]{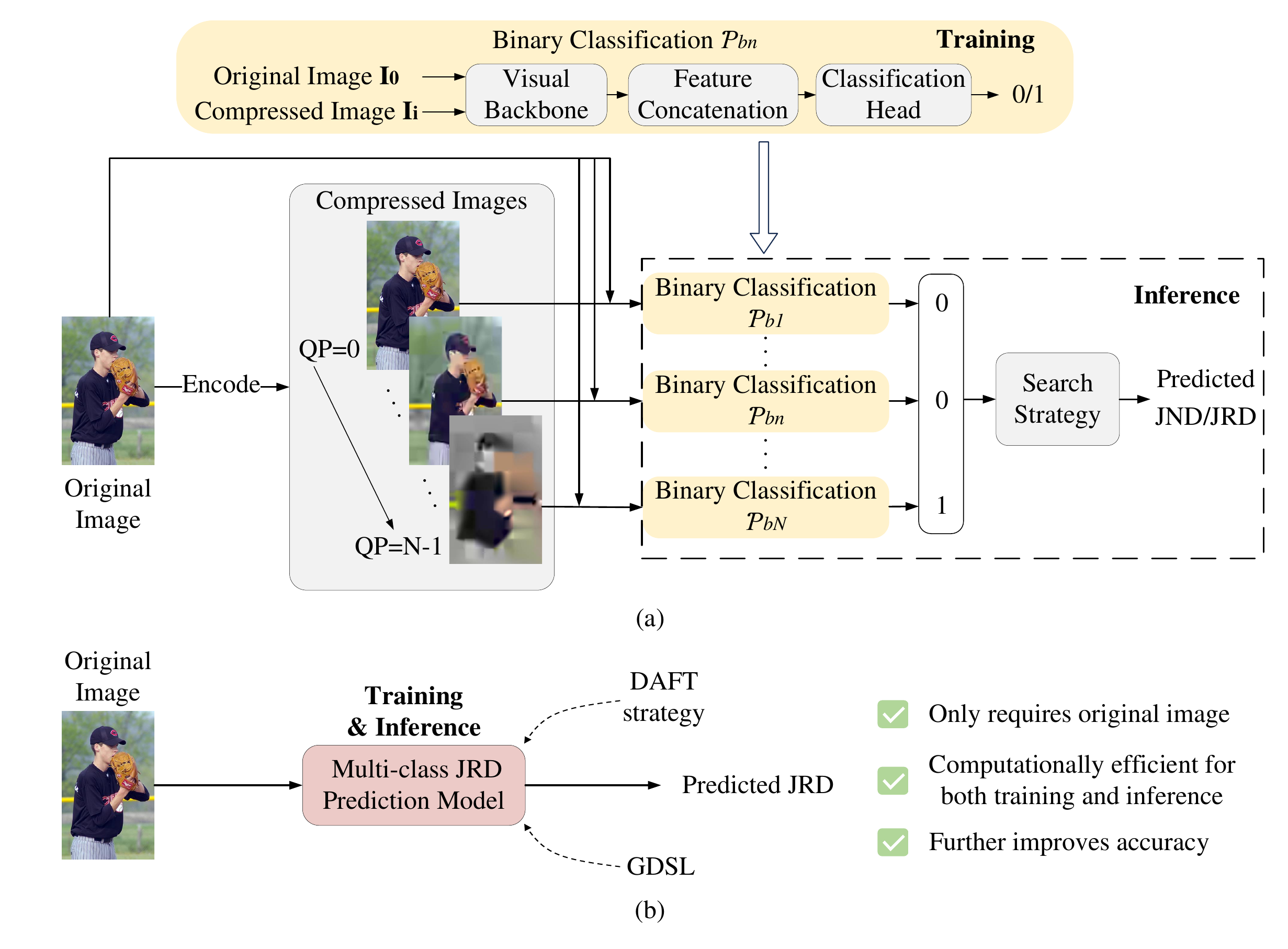}
        \captionsetup{justification=justified}
        \caption{Framework of the JRD prediction models, (a) Binary-class based JRD prediction\cite{zhang2021just,zhang2023learning}, 
 (b) Multi-class based JRD prediction.}
        \label{formulation}
\end{figure}

\subsection{Motivations and Problem Formulation}
\label{section:motivation}
Studying of the JRD mechanism and developing JRD prediction model has drawn much attention due to its promising applications in VCM. Predicting JRD is to predict the distortion threshold that affects machine vision performance among all candidate distortion levels, which can be formulated as a multi-class classification problem and solved with learning-based prediction. The JRD $\mathcal{\hat{J}}_{i,j}$ is predicted as
\begin{equation}
    \mathcal{\hat{J}}_{i,j}=\mathbb{P}_{N}(\mathbf{I}_{i,j})
    \label{eq0}
\end{equation}
where $\mathbf{I}_{i,j}$ is the $i$-th object in $j$-th input image and $\mathbb{P}_{N}$ is a learned $N$-class JRD prediction model. However, there are 100 quality scales in JPEG and 51/63 scales in HEVC/VVC. Due to the large number of scales and their fine-grained distortion levels, it is challenging to predict the JRD value accurately.

To handle this problem and predict the JRD accurately, the existing JRD prediction models \cite{zhang2021just,zhang2023learning} decomposed the multi-class classification as multiple binary classifications and proposed to generate additional distorted images as input for comparison with the source pristine image. The framework of these JRD prediction models, as shown in Fig. \ref{formulation} (a), is formulated as
\begin{equation}
\mathcal{\hat{J}}_{i,j}=\mathbb{S}(\mathbb{P}_{b_1}(\mathbf{I}_{o}, \mathbf{I}^{1}_{i,j}),\mathbb{P}_{b_2}(\mathbf{I}_{o}, \mathbf{I}^{2}_{i,j}),\dots, \mathbb{P}_{b_N}(\mathbf{I}_{o}, \mathbf{I}^{N}_{i,j}))
\label{eq1}
\end{equation}
where $\mathbf{I}_o$ is source image and $\mathbf{I}^{n}_{i,j}$, $n\in[1...N]$, is distorted image at level $n$, $\mathbb{P}_{b_n}$ is the $n$-th binary classification, $\mathbb{S}()$ is a post-processing search strategy for error correction. In these works, $N$ distorted images shall be generated and $\mathbb{P}_{b_n}$ shall be run $N$ times, which causes additional, unaffordable computational overhead as $N$ becomes large. Secondly, the multiple binary classifiers make decisions independently of each other, which may cause contradiction and false prediction.
Finally, the prediction accuracy could be further improved.

To address these problems, we take advantage of powerful deep learning in solving the multi-class prediction problem. Due to the powerful capability of deep vision transformer in computational intelligence, we propose to exploit it to use single image input and solve the problem in Eq. \ref{eq0} directly.

\subsection{DT-JRD Prediction Framework and Architecture}
\label{section:framework}
We formulate the JRD prediction problem as a multi-classification problem with $N$ classes, as shown in Fig. \ref{formulation} (b). Firstly, only the original image is required in predicting the JRD in Eq. \ref{eq0}. Secondly, the prediction model only required to be performed only once. Thirdly, we proposed to use Vision Transformer (ViT)\cite{dosovitskiy2020image} as the backbone to exploit its powerful attention-based long-range dependency modeling and effective feature extraction capabilities in the JRD prediction, where a novel training strategy is presented. The architecture of the proposed DT-JRD model, as shown in Fig. \ref{fig3.1}, where key modules include an improved embedding, a content and distortion feature extraction, and a multi-class classification. Moreover, the DT-JRD is guided by the GDSL-based loss function, which augments training labels and improves the JRD prediction accuracy.

\begin{figure}[t]
	\centering
	\includegraphics[width=0.5\textwidth]{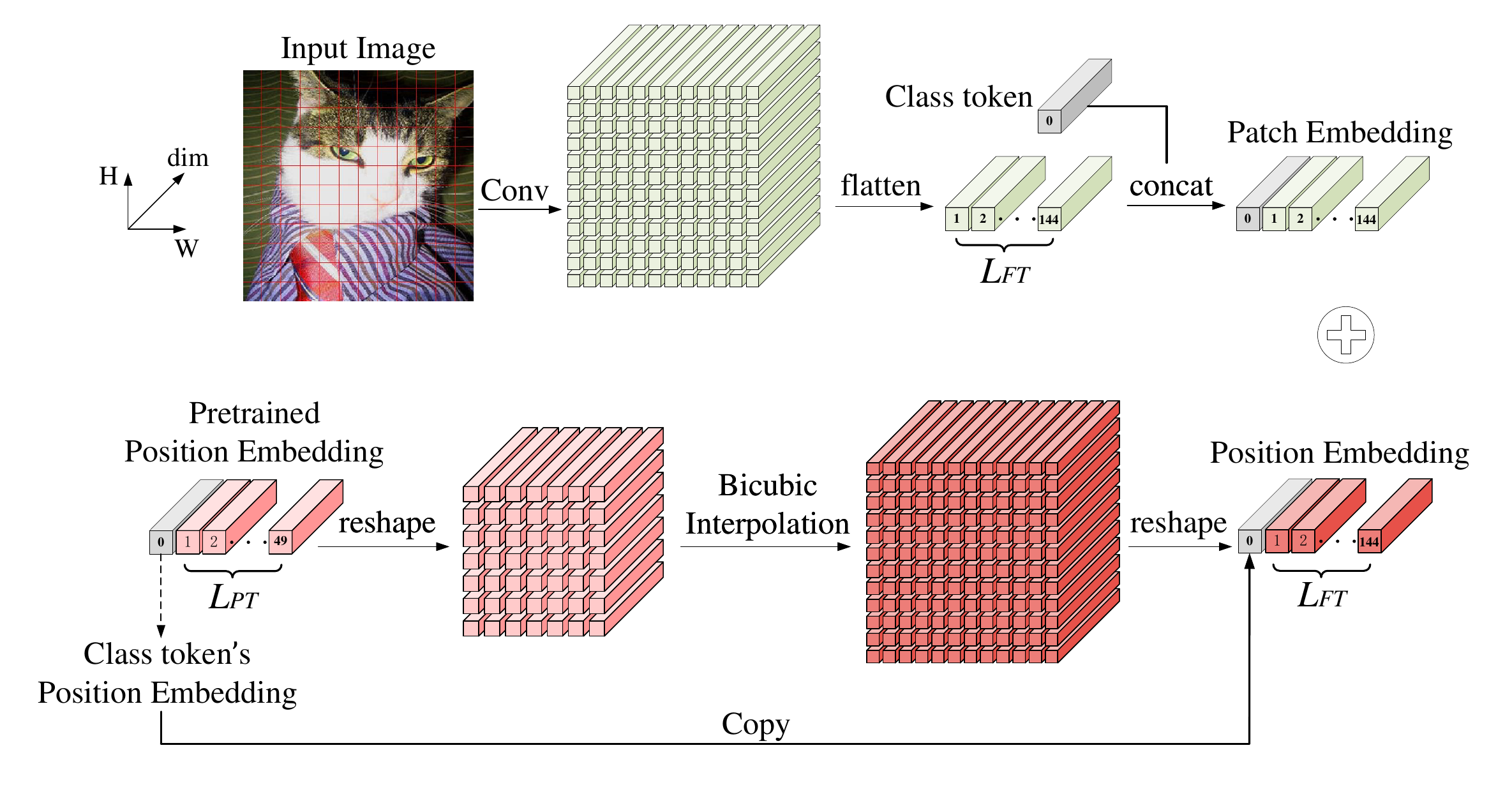}
        \captionsetup{justification=justified}
	\caption{2D Interpolation for Position Embedding.}
	\label{interpolation}
\end{figure}

\begin{figure*}[t]
	\centering
    \captionsetup{justification=justified}
	\includegraphics[width=0.85\textwidth]{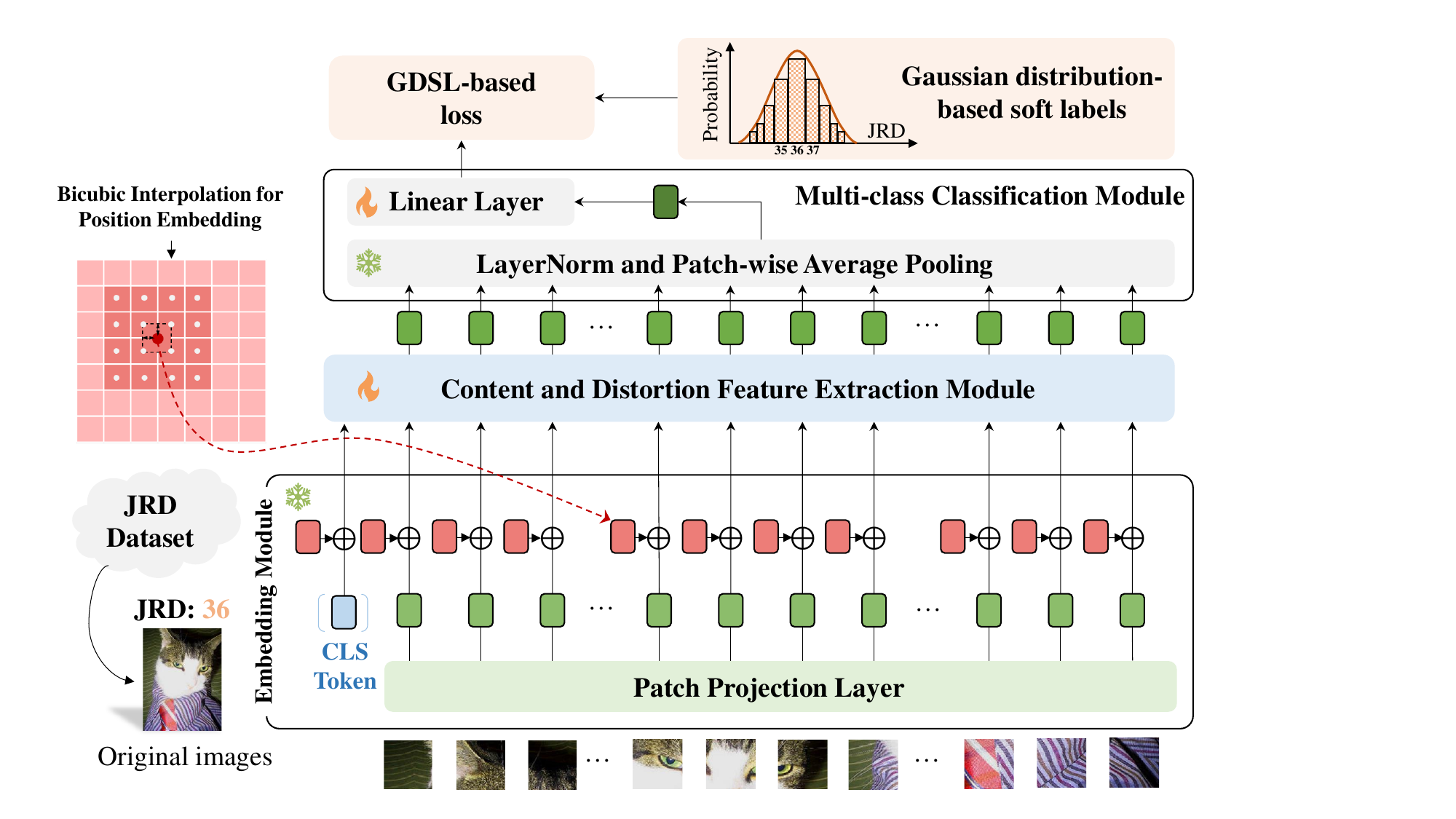}
	\caption{The framework of the proposed DT-JRD prediction model.}
	\label{fig3.1}
\end{figure*}

\subsubsection{{2D Interpolation for Position Embedding}}

In this paper, we use the JRD dataset in \cite{zhang2023learning}. On the one hand, since the images of this dataset are cropped from the COCO\cite{lin2014microsoft} dataset images based on the objects, their sizes vary with each other. Small, medium, and large size objects account for 12\%, 35\%, and 53\% in the JRD dataset, respectively.
For batch training, it is helpful to use a uniform input image size that preserves the pixel information of the images with an affordable computational burden.
On the other hand, in the ViT, the input image size determines the length of the patch sequence and their embedding positions. If a constant patch size is used (e.g. 32$\times$32) when fine-tuning a higher resolution image, it will generate new locations to be encoded, so the pre-trained position embedding may no longer be meaningful. 
One solution is to directly use fixed position encoding of the sine and cosine functions, but it does not inherit the prior knowledge of the pre-trained model.
Therefore, it is necessary to develop 2D interpolation on the pre-trained position embedding according to their location, which enhances the flexibility and compatibility of using different input image sizes.

Fine-tuning with a higher image resolution than that used in the pre-trained model can achieve better performance in downstream tasks\cite{kolesnikov2020big,touvron2019fixing}.
Therefore, to improve the prediction accuracy, we use 384$\times$384 image on ViT for fine-tuning, which is a higher resolution than that of pre-training. Then, we apply a 2D interpolation on the pre-trained position embedding weights. 
Fig. \ref{interpolation} shows the flowchart of the 2D interpolation for position embedding, where dim is the dimension of embeddings, $L_{PT}$ and $L_{FT}$ represent the sequence length in the pre-training and fine-tuning stage, respectively. Features from the convolutions of the input image are flattened, which is then concatenated with a class token for patch embedding. Then, pre-trained position embedding is reshaped and interpolated to align with the higher-resolution patch embedding. Bicubic interpolation is used to provide smoother and more accurate interpolation results while reducing edge effect. Concretely, $L_{PT}=224\times224/(32\times32)=49$, $L_{FT}=384\times384/(32\times32)=144$.
This sequence length contains enough pixel information for prediction without adding much computational burden.
Thus, patch embeddings and position embeddings have a one-to-one correspondence. Since the patch embeddings are obtained from divided image blocks, there is a spatial correlation among them. Therefore, performing 2D interpolation on the pre-trained position embedding weights is beneficial to enable the position embedding to inherit this spatial correlation.

\subsubsection{{Content and Distortion Feature Extraction}}
The JRD dataset contains more than 29,000 object images with over 80 content categories and each image has 64 distortion categories. To handle the diverse image contents and distortion, a content and distortion feature extraction module is developed based on the transformer encoder block from ViT. It consists of a stack of 24 Transformer encoder blocks, each of which is composed of multi-head attention, layer normalization, Multi-Layer Perceptron (MLP), and residual connection, as shown in Fig. \ref{fig3.1}. Distortion usually manifests in the frequency domain as a specific frequency shift or noise. Multi-head attention enhances feature interactions for different frequency components, thus providing the DT-JRD model with an exceptional perception capability of global information in capturing content and distortion features. Therefore, semantic JRD features are effectively extracted through the stacked encoder blocks.

\subsubsection{{Multi-class Classification for JRD Prediction}}
After the content and distortion feature extraction module, features are aggregated by means of patch-wise average pooling and the predicted probabilities for each category are then output through the classification head. The output features of the Transformer encoder contain a class token and multiple patch tokens. In the original ViT model, only a class token is extracted for classification head prediction. The class token of the pre-trained ViT model contains more prior information related to the image content categories. However, the JRD prediction model expects the final extracted features to incorporate more semantic information related to the image JRD. Hence, a patch-wise average pooling is used to pool the patch tokens from the Transformer encoder and get $\hat{z}$, which is
\begin{equation}
   {\hat{z}} = \frac{\sum_{i=1}^{L}\phi_{LN}(\mathbf{z}^i)}{L}
   \label{GAP}
\end{equation}
where $\mathbf{z}^i$ refers to the $i$-th patch token, $L$ refers to sequence length of fine-tuning $L_{FT}$ and $\phi_{LN}$ refers to layer norm. Finally, we obtain the predicted probability $p$ through a linear layer for classification, as shown in Fig. \ref{fig3.1}.

\subsection{Distortion-aware Learning Strategy}
\label{section:learning strategy}
\begin{figure}[t]
	\centering
    \includegraphics[width=0.48\textwidth]{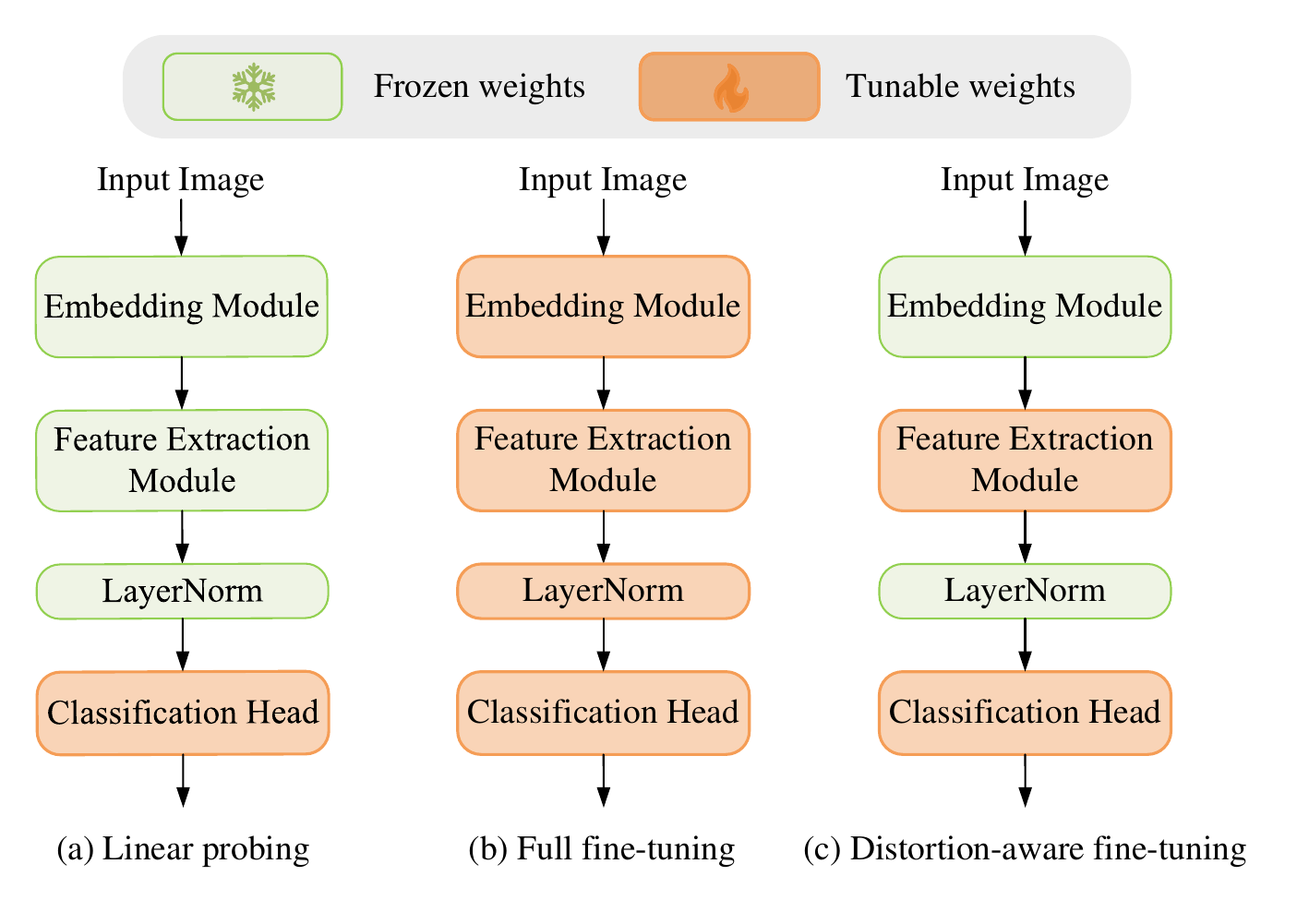}
    \captionsetup{justification=justified}
    \caption{Strategies for learning the DT-JRD model.}
    \label{fig4}
\end{figure}

In DT-JRD, training from scratch requires a large number of labeled data samples. The JRD dataset containing a total of 29218 object images and corresponding JRD labels is rather small. To handle this problem, we propose a distortion-aware learning strategy for DT-JRD, which reduces the number of training parameters while maintaining the model learning ability.
Fig. \ref{fig4} shows a comparison between conventional and the proposed learning strategies.
Fig. \ref{fig4} (a) shows the Linear Probing (LP), where only the weights of the classification head are updated. The LP strategy only updates the classification parameters, which is parameter-efficient but not effective enough for predicting JRD because the frozen ViT backbone network mainly extracts image content features without considering distortion features.
On the other hand, the Full Fine-tuning (FF) strategy in Fig. \ref{fig4} (b) utilizes all the learnable parameters of the large-scale pre-trained models, but performing it on smaller datasets like the JRD dataset may cause over-fitting. Also, FF may cause oblivion that loses some useful knowledge.

To handle these problems, we propose the Distortion-Aware Fine-Tuning (DAFT) strategy in Fig. \ref{fig4} (c) to strike a balance between being task-specific and being task-agnostic.
First, freezing the shallow layers is preferable to training all the parameters\cite{chen2021empirical}. The shallow layers extract low-level and general features, while the deep layers extract high-level semantic features and have a higher impact on the final classification. Therefore, freezing the weights of the shallow layers like patch embedding, class token, and position embedding is beneficial to protect the low-level feature.
Furthermore, freezing the final layer normalization module helps to maintain the stable feature representation learned from pre-training while reducing the risk of over-fitting and avoiding the problem caused by too sensitive learning rate in parameter updates.
Conversely, feature extraction module and classification head are trained to improve representation learning and prediction capability. Based on this fine-grained DAFT strategy, we apply it to train DT-JRD, as shown in Fig. \ref{fig3.1}, where blocks with the symbol ``fire" are re-trainable and blocks with the symbol ``snow" are frozen. The proposed DAFT strategy not only utilizes the rich feature representation of the pre-trained model but also enables a robust JRD prediction.

\subsection{GDSL based Loss Function}
\label{section:GDSL}
In traditional image classification, a classification model aims to establish a relationship between image content and categories, where the performance is typically measured with the ratio of images that can be correctly classified as the ground truth label, i.e., classification accuracy. Fig. \ref{fig5} shows the probability assignment of different types of labels for classification. Fig. \ref{fig5} (a) shows the one-hot label for conventional image classification, where the probability is 1 at the ground-truth JRD and 0 for the others. If the JRD is not correctly predicted, 100\% penalty will be given to the negative log-likelihood of the predicted probability for the true class. Fig. \ref{fig5} (b) shows a smooth label where a larger probability $\epsilon$ (e.g. 0.9) is assigned to the true class and the remaining probability $1-\epsilon$ is evenly distributed among the other classes, preventing the model from predicting a category too confidently. This smooth label helps to enhance generalization ability and relieve over-fitting problem \cite{szegedy2016rethinking}. However, all classes except the true class are equally important in calculating loss, which does not reflect the differences in penalties for different positions. 

\begin{figure}[t]
    \centering
        \subfloat[]{
        \includegraphics[width=0.23\textwidth]{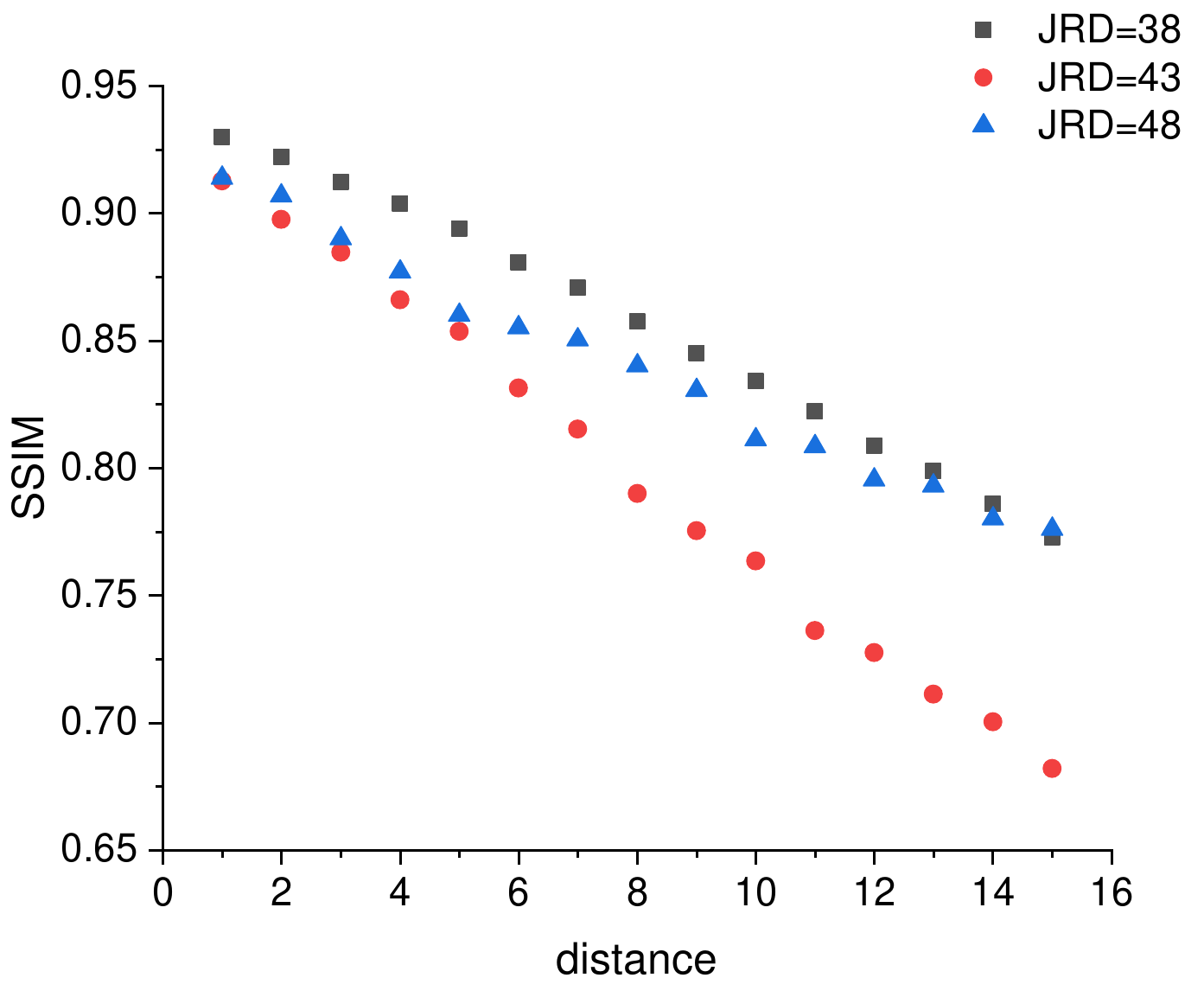}
        \label{fig:sub5.1}}
        \subfloat[]{
        \includegraphics[width=0.23\textwidth]{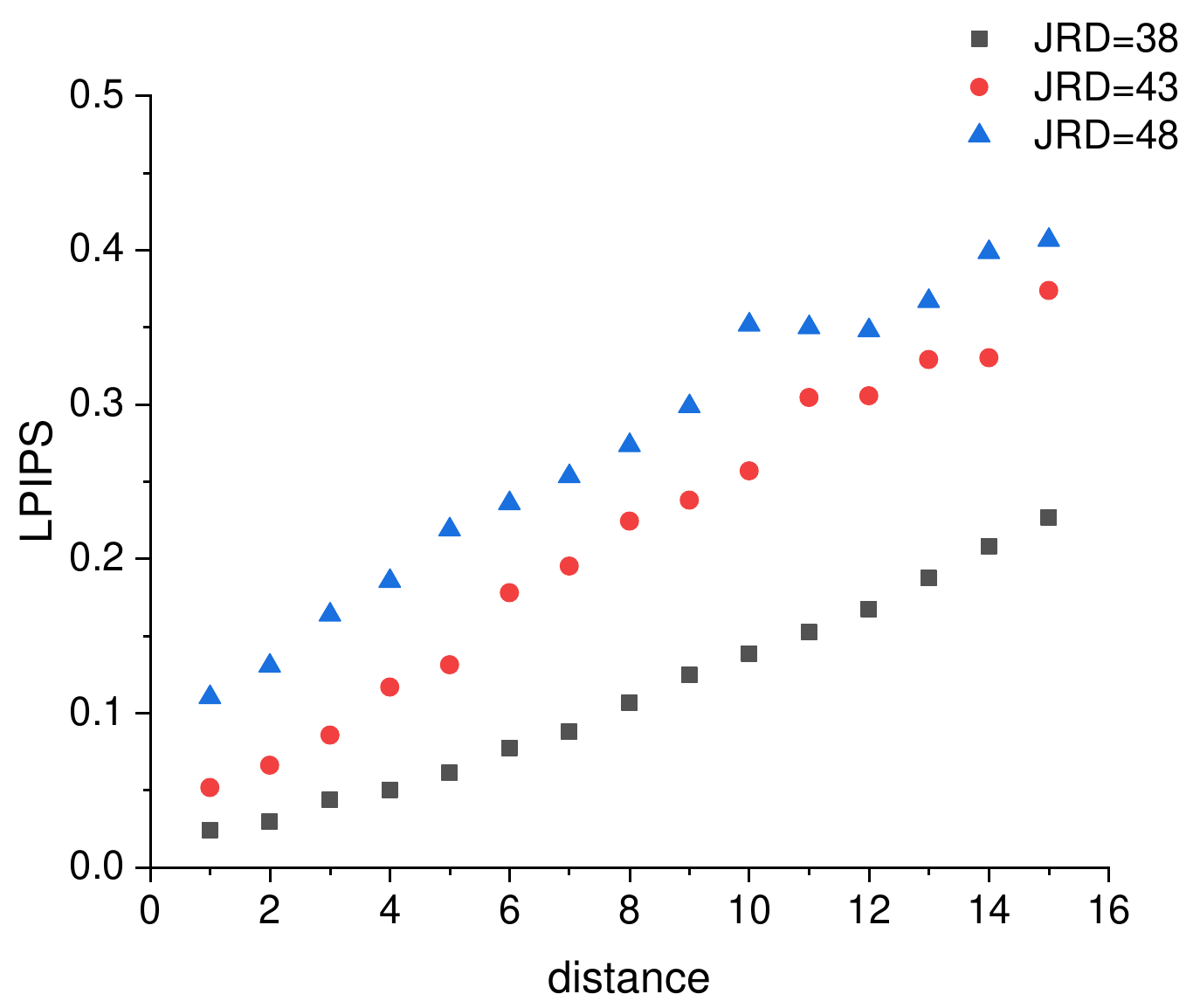}
        \label{fig:sub5.2}}
    \captionsetup{justification=justified, singlelinecheck=false}
    \caption{Relationship between the distance to the JRD and the similarity to the JRD image for three samples with JRD values of 38, 43, and 48.}
	\label{scatter}
\end{figure}


Different from conventional image classification, the class labels in JRD prediction task have spatial correlation. Although the ground truth JRD $\mathcal{J}_{i,j}$ is the key objective, the neighboring JRDs, such as $\mathcal{J}_{i,j}+k$ and $\mathcal{J}_{i,j}-k$, can also be regarded as high confidence labels when $k$ is small. When $k$ is large, the confidence could be reduced. 
To validate this assumption, we performed a statistical analysis on the correlation between the ground truth JRD image and its neighboring images. Three sample images with JRD values of 38, 43, and 48 were analyzed, $k\in[1,15]$. Structural Similarity Index Measure (SSIM)\cite{SSIM} and  Learned Perceptual Image Patch Similarity (LPIPS)\cite{LPIPS} were used to measure the similarity between the JRD image and images with adjacent distortion levels. High SSIM values and low LPIPS values indicate high similarity between the two images.

Fig. \ref{scatter} shows the relationship between the distance $|k|$ and the similarity between images of JRD $\mathcal{J}_{i,j}$ and $\mathcal{J}_{i,j}+k$. We can observe that there is a strong similarity between the JRD images and images with adjacent distortion levels. Furthermore, a continuous decreasing trend can be noted as the absolute distance from JRD $|k|$ increases. 
Moreover, the JRD labeling process \cite{zhang2023learning} revealed that some fluctuating points occur near the perceptual threshold, reflecting a similar response of machine vision in the JRD neighborhood. Therefore, the neighboring labels of the ground truth, which have high similarity, are acceptable to be used as augmented data for model training. 
Based on the above analysis, we proposed the GDSL-based loss function for training the DT-JRD model.

\begin{figure*}[t]
	\centering
        \captionsetup{justification=justified}
        \includegraphics[width=0.95\textwidth]{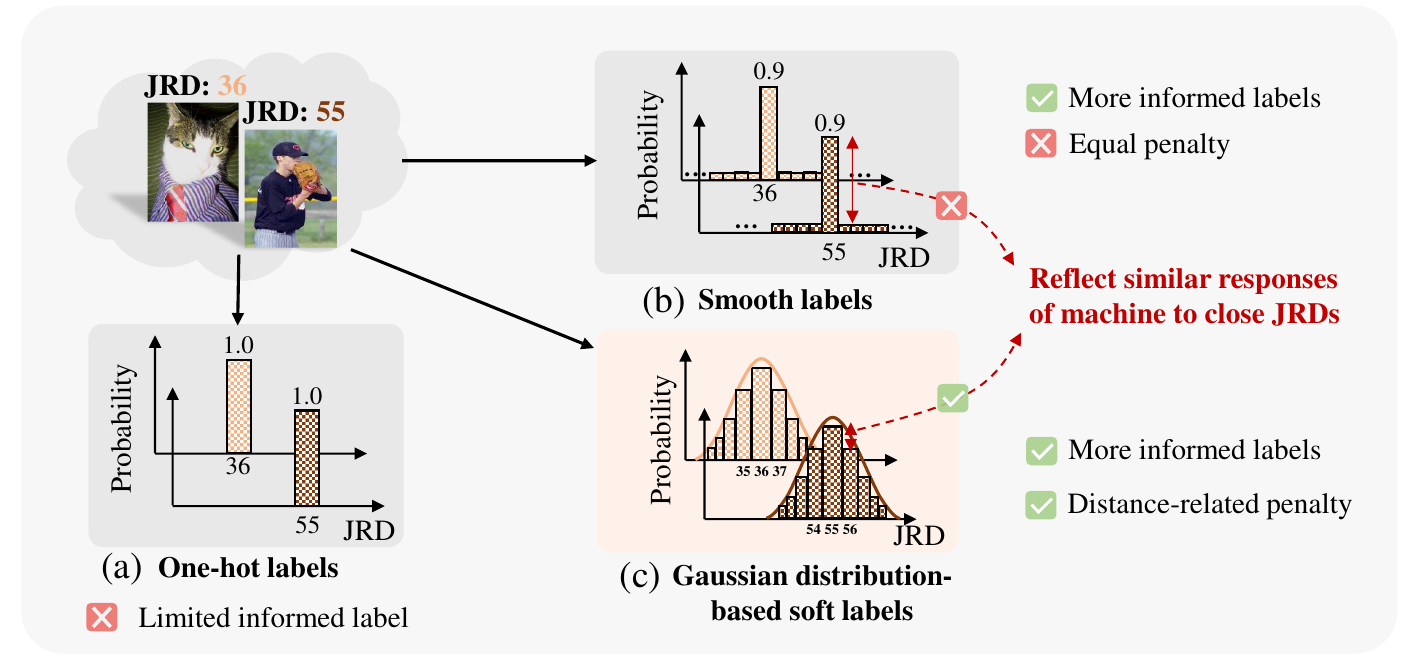}
        \caption{Visualization for different types of labels and their respective advantages and limitations.}
        \label{fig5}
\end{figure*}

As shown in Fig. \ref{fig5} (c), GDSL fits the probability distribution on each category as a Gaussian function, where higher probability is given to the labels closer to the ground truth class, and lower probability is given to the distant labels. 
For an object with JRD $\mu$, we assume the probability $q(x)$ of soft-labeled JRD obeys a Gaussian distribution with a mean of $\mu$ and a standard deviation of $\sigma$, which is calculated as
\begin{equation}
\label{eq_1}
q_N(x;\mu,\sigma)=\frac{\frac{1}{\sqrt{2 \pi} \sigma} e^{-\frac{(x-\mu)^2}{2 \sigma^2}}} {\sum\limits_{x=0}^{N-1} \frac{1}{\sqrt{2 \pi} \sigma} e^{-\frac{(x-\mu)^2}{2 \sigma^2}}},
\end{equation}
where $x$ is the QP, i.e. JRD candidate, $x \in\{0,1,2, \ldots, N-1\}$, $N$ is 64 for VVC in this paper. Therefore, we propose a GDSL-based loss function to train the DT-JRD model. The optimal parameters $\theta^*$ is learned as
\begin{equation}
\begin{cases}
   \theta^* = \operatorname*{arg\,min}\limits_{\theta}{\mathcal{L}_{GDSL}}\\
   \mathcal{L}_{GDSL}=-\sum\limits_{x=0}^{N-1} {L}_{N}(x;\mu,\sigma)\log{\mathbb{P}_{N(\mathbf{I}_{o};\theta)}(x)}   
\end{cases},
\label{argmin}
\end{equation}
where ${L}_{N}(x)$ are the soft labels of JRD, i.e. ${L}_{N}(x;\mu,\sigma)=q_N(x;\mu,\sigma)$, $\mathbb{P}_{N}$ is the probability output from the $N$-class DT-JRD prediction model and $\mathbf{I}_{o}$ is the source image. In the case of the smooth label, ${L}_{N}(x)$ is either $\epsilon$ for the JRD or $1-\epsilon$ for the rest. The $\mathcal{L}_{GDSL}$ has two advantages as compared with the one-hot and smooth labels. Firstly, more labels are available and can be used for model training, which is equivalent to data label augmentation. Secondly, it reflects the distance penalty from the ground truth JRD.

\begin{table}[t]
	\centering
    \caption{Performance validation of DT-JRD using different types of labels}
	\label{tab_val}
	\setlength{\tabcolsep}{8mm}
        {
		\begin{tabular}{cc}
			\bottomrule[1.5pt]
			Type of labels  & MAE \\
			\hline
			one-hot label & 5.496 \\
            smooth label($\epsilon$=0.9) & 5.599 \\
			GDSL($\sigma$=2) & 5.374	\\
			GDSL($\sigma$=3) & \textbf{5.317} \\
			GDSL($\sigma$=4) & 5.335 \\
			GDSL($\sigma$=5) & 5.382 \\
			GDSL($\sigma$=6) & 5.394	\\
			GDSL($\sigma$=7) & 5.421	\\
			\toprule[1.5pt]
		\end{tabular}
	}
\end{table}

To validate the performance of the GDSL-based loss function, JRD prediction is validated on the validation set. To verify the effectiveness of $\mathcal{L}_{GDSL}$, we conducted experiments on the validation set to compare the performance of one-hot label, smooth label, and GDSL-based loss with different $\sigma \in \{2, 3, 4, 5, 6, 7\}$. The JRD dataset was randomly divided into training, validation, and test sets in an 8:1:1 ratio. Table \ref{tab_val} shows the JRD prediction accuracy measured with Mean Absolute Error (MAE) from the DT-JRD that was trained from different labels. The first two rows show the MAEs of the DT-JRD model trained with one-hot and smooth labels, which are 5.496 and 5.599 on the validation set, respectively. The bottom six rows show the MAEs from GDSL using different $\sigma$. We can observe that the MAE ranges from 5.317 to 5.421, which are lower than those of the one-hot and smooth labels. The MAE achieves the minimum value of 5.317 when $\sigma$ is 3, which is selected. 

\begin{figure*}[t]
	\centering
        \captionsetup{justification=justified}
	\includegraphics[width=0.85\textwidth]{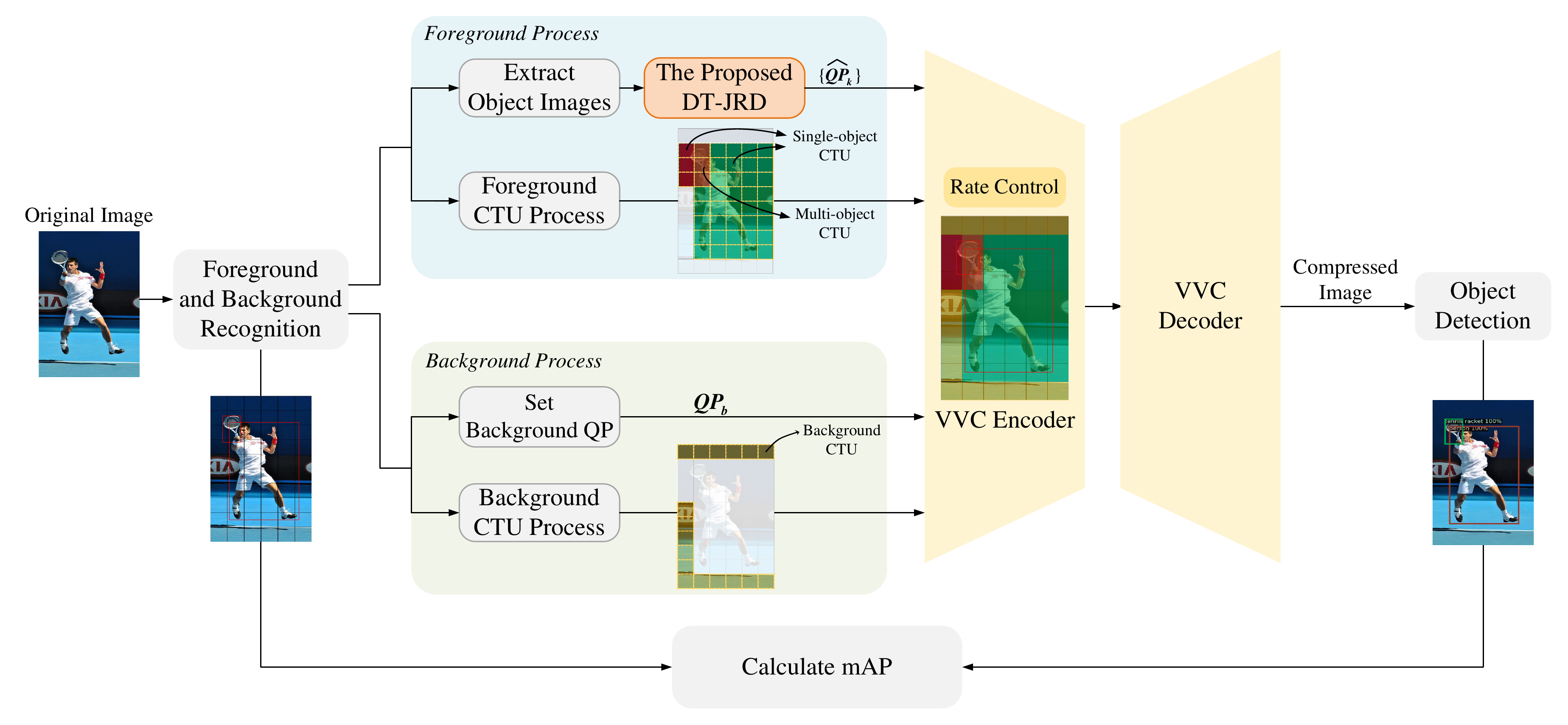}
	\caption{The processing workflow for DT-JRD based VCM optimization.}
	\label{coding method}
\end{figure*}

\subsection{DT-JRD based VCM Optimization}
\label{section:coding}
 The predicted JRD can then be applied to control the coding quality of the corresponding target so that the compressed image can satisfy the characteristic of JRD at the optimal bit rate. Fig. \ref{coding method} shows the flowchart of the DT-JRD based VCM optimization, where the predicted JRD is applied to guide the encoding process of the object.
Blocks located to get the bounding box are regarded as objects, denoted as $\{\mathbf{B}_k=\left[x_{ul}, y_{ul}, x_{lr}, y_{lr}\right]\}$, where $k \in\{1,2, \ldots, M\}$, $M$ is the number of objects detected in an image, $\left(x_{ul}, y_{ul}\right)$ and $\left(x_{lr}, y_{lr}\right)$ represent the upper-left and bottom-right coordinates of the object bounding box, respectively.
Then, the object images are cropped and fed into the proposed DT-JRD to get their predicted JRD $\{\widehat{QP}_{k}\}$, which was used to guide the VCM optimization. The images are converted to YUV420 and divided into 64$\times$64 Coding Tree Units (CTUs), which are classified into two key categories, i.e., object CTU and background CTU. One CTU is regarded as an object CTU if it has an intersection with the object. The object CTU is encoded with the predicted JRD $\{\widehat{QP}_{k}\}$ or below to maintain high performance of machine vision task. If the CTU belongs to multiple objects, the minimum JRD among the predicted JRDs $\{\widehat{QP}_{k}\}$ for different objects was selected as the coding QP. The background CTU is encoded with background QP $QP_b$, which is larger than $\{\widehat{QP}_{k}\}$ for saving bit rate. The object and background CTUs are then encoded with the VVC encoder for transmission. At the decoder side, the reconstructed images from VVC decoder are converted to RGB, and then input to Faster-RCNN \cite{ren2015faster} for object detection performance evaluation.

\section{Experimental Results and Analysis}
\label{section:experiment}
In this section, we experimentally validate the proposed DT-JRD prediction model and DT-JRD based VCM.
First, the experimental settings are presented. Second, the DT-JRD prediction model is trained, validated, and compared with two state-of-the-art JRD prediction models. Third, ablation studies on DT-JRD are performed. Finally, coding experiments are performed to validate the performance of DT-JRD based VCM.

\subsection{Experimental Settings}
\subsubsection{Dataset and Model Configurations}
The JRD dataset \cite{q5fq-d638-23} is randomly divided into training, validation, and test sets in an 8:1:1 ratio. It is noted that there are some special samples in the dataset where large objects contain small objects. To avoid data leakage and to guarantee the isolation of the test set from the training set, we divide all objects on the same original image into the same subset. To facilitate batch training, all object images are re-scaled by bilinear interpolation to the same size (384$\times$384). Beyond that, the training images are horizontally flipped with a probability of 0.5 to increase the data diversity during training and help the model learn more generalized features. Our DT-JRD prediction model adopted ViT-Large/32 pre-trained on ImageNet-21k. We trained the DT-JRD model on 2 RTX3090 graphics cards. The batch size was set as 32, and the learning rate was initially set to 0.01 and adjusted by the cosine decay strategy. Momentum-based stochastic gradient descent method was chosen to be the optimizer, with momentum value set to 0.9 and weight decay to $5\times10^{-5}$. The proposed GDSL-based loss was used.

To verify the effectiveness of the DT-JRD, we compared it with other two state-of-the-art JRD prediction schemes which use ensemble learning or a binary classifier, denoted as ``EL-JRD''\cite{zhang2021just} and ``BC-JRD''\cite{zhang2023learning} in Table \ref{table3.3}. We implemented and retrained EL-JRD with eight binary classifiers by using VGG19 as the backbone for one against all classification, while BC-JRD was trained with only one binary classifier to perform classification multiple times.

\subsubsection{Evaluation Metrics}
To evaluate the prediction performance of the proposed DT-JRD model, we use MAE between the predicted JRD ($\hat{\mathcal{J}}_{i, j}$) and the ground truth JRD ($\mathcal{J}_{i, j}$) of the $i$-th object in the $j$-th image, which is calculated as\cite{zhang2023learning}
\begin{equation}
\label{eq_6}
E_A=\frac{1}{n} \sum_{j=1}^n\left(\frac{1}{m_j} \sum_{i=1}^{m_j}\left|\hat{\mathcal{J}}_{i, j}-\mathcal{J}_{i, j}\right|\right),
\end{equation}
where $n$ is the total number of images in the JRD dataset, and $m_j$ is the number of objects in the $j$-th image.
Since JRD of more than 80\% objects are located in the range of [27, 51], MAE of the majority $E_{[27,51]}$ is additionally calculated as\cite{zhang2023learning}
\begin{equation}
\label{eq_7}
\left\{\begin{array}{l}
E_{[27, 51]}=\frac{1}{\sum_{k=27}^{51} n_k} \sum_{k=27}^{51}(E(k)) \\
E(k)=\sum_{i=1}^{n_k}\left|\hat{\mathcal{J}}_{i}-k\right|
\end{array}, \right.
\end{equation}
where $n_k$ and $\hat{\mathcal{J}}_{i}$ respectively represent the number of objects and the predicted JRD with ground truth equal to $k$, $E(k)$ is the sum of the predicted error whose ground truth JRD is $k$.

In addition, the JRD prediction performance can be evaluated in terms of $\lvert\Delta$PSNR$\rvert$, $\lvert\Delta$SSIM$\rvert$\cite{SSIM}, $\lvert\Delta$B$\rvert$ and $R^2$. $\lvert\Delta$PSNR$\rvert$ calculates the PSNR MAE between the image compressed by the ground truth JRD and the one compressed by predicted JRD. $\lvert\Delta$SSIM$\rvert$ and $\lvert\Delta$B$\rvert$ are calculated in the same way for SSIM and bit rate.
$R^2$ of PSNR refers to the correlation between predicted JRD in PSNR and ground truth JRD in PSNR.

\subsubsection{Coding Settings and Evaluations}
We instantiated the proposed DT-JRD based VCM method on the JRD testing set with VVC. 
For fair comparison, the ground truth JRD, and the predicted JRD from two benchmarks EL-JRD and BC-JRD were also applied to VCM. VVC configured with all-intra was used as the base encoder and $QP$ $\in$ [25, 27, 29, 31, 33].
Considering the rate-accuracy balance in practical applications, it is feasible to encode the predicted JRD by adding an offset. We denote the offset as $\Delta QP$, where $\Delta QP$ $\in$ [-4, -3, -2, -1, 0].

To evaluate the efficiency of VCM, we use the rate-accuracy curve to represent the relationship between machine vision performance and bit rate. The detection performance of the machine vision is typically measured by mean Average Precision (mAP) calculated at IOU thresholds of 50$\%$ and 75$\%$, denoted as ``mAP@0.5" and ``mAP@0.75", respectively. The detected results of the original images by using the Faster R-CNN\cite{ren2015faster} are used as ground truth in calculating the mAP. Other detailed settings can be referred to in\cite{zhang2023learning}.

\subsection{JRD Prediction Performance}
In this section, we comprehensively analyzed the prediction performance of the proposed DT-JRD prediction model and compared it with ``EL-JRD''\cite{zhang2021just} and ``BC-JRD''\cite{zhang2023learning} in Table \ref{table3.3}. Specifically, objects were divided into small, medium, and large sizes based on the criteria of the COCO dataset. In terms of content categories, person and car, which are the two most frequently sampled categories, were selected for illustration. $E_A$ in the first five columns of the table refers to MAE for that category, while $E_A$ and $E_{[27,51]}$ in the last two columns of the table are calculated on the whole test set and within a range of frequently used QPs, respectively.

\setlength{\tabcolsep}{4pt} 
\begin{table}[t]
	\centering
	\caption{JRD Prediction Performance Comparison on Different Sizes and Categories.}
	\begin{tabular}{cccccccc}
		\bottomrule[1.5pt]
		\multirow{2}[4]{*}{Model} & \multicolumn{5}{c}{$E_A$}               & \multicolumn{1}{c}{\multirow{2}[4]{*}{$E_A$}} & \multicolumn{1}{c}{\multirow{2}[4]{*}{$E_{[27,51]}$}} \bigstrut\\
		\cline{2-6}          & Small   & Medium   & Large   & Person & Car  &       &  \bigstrut\\
		\hline
  		EL-JRD\cite{zhang2021just} & 9.47 & 8.57  & 7.99  & 6.80  & 10.19 & 8.37  & 7.71 \bigstrut[t]\\
		BC-JRD\cite{zhang2023learning} & 6.24  & 6.21  & 6.58  & 5.07  & 6.56  & 6.41  & 4.90 \\
		DT-JRD  & \textbf{5.61} & \textbf{5.50} & \textbf{5.62} & \textbf{4.39}  & \textbf{5.30} & \textbf{5.57} & \textbf{4.20} \bigstrut[b]\\
		\toprule[1.5pt]
	\end{tabular}
	\label{table3.3}
\end{table}%

As shown in Table \ref{table3.3}, the proposed DT-JRD method is meticulously compared with other approaches in terms of performance across various object sizes and categories. It can be observed that the $E_A$ and $E_{[27,51]}$ achieved by the EL-JRD are 8.37 and 7.71, which are the highest among the three schemes. In comparison, the BC-JRD achieves $E_A$ and $E_{[27,51]}$ values of 6.41 and 4.90, demonstrating a notable improvement over the EL-JRD. Our proposed DT-JRD model, on the other hand, consistently outperforms the other methods across diverse object sizes and categories. Specifically, it achieves 5.57 and 4.20 of $E_A$ and $E_{[27,51]}$, respectively, representing enhancements of 33.45\% and 45.53\% over the EL-JRD and improvements of 13.10\% and 14.29\% compared to the BC-JRD. Furthermore, DT-JRD shows superior performance for small, medium, and large objects, with $E_A$ values of 5.61, 5.50, and 5.62, respectively. These results outperform those of the best existing method, BC-JRD, whose values are 6.24, 6.21, and 6.58, by 10.10\%, 11.43\%, and 14.59\%. In addition, DT-JRD maintains superior performance on representative categories, namely ``person" and ``car". Importantly, the proposed DT-JRD model only requires a single source image for JRD prediction, which reduces computational complexity and enhances the flexibility.

\begin{figure*}[t]
    \centering
        \subfloat[EL-JRD]{
        \includegraphics[width=0.32\textwidth]{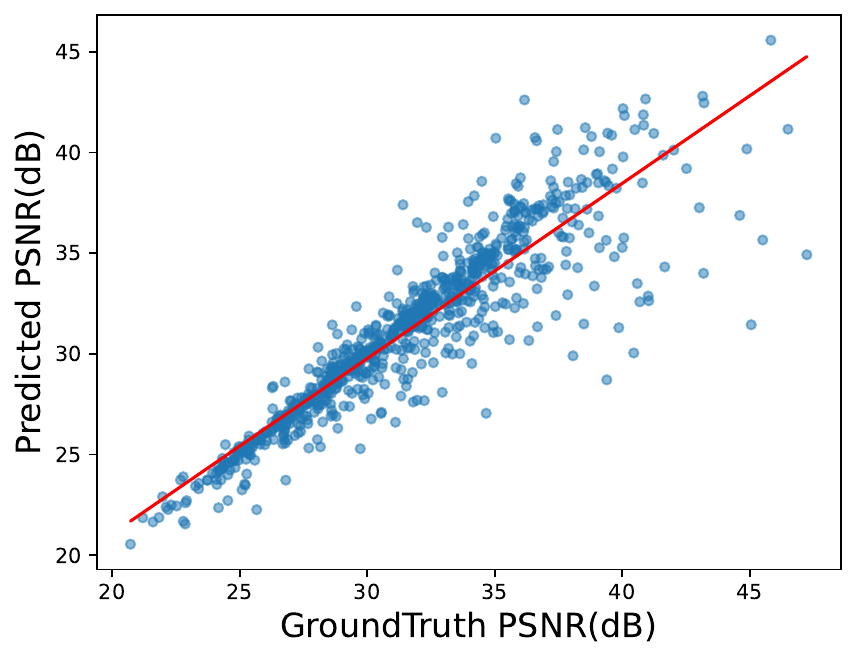}
        \label{fig:sub8.1}}
        \subfloat[BC-JRD]{
        \includegraphics[width=0.32\textwidth]{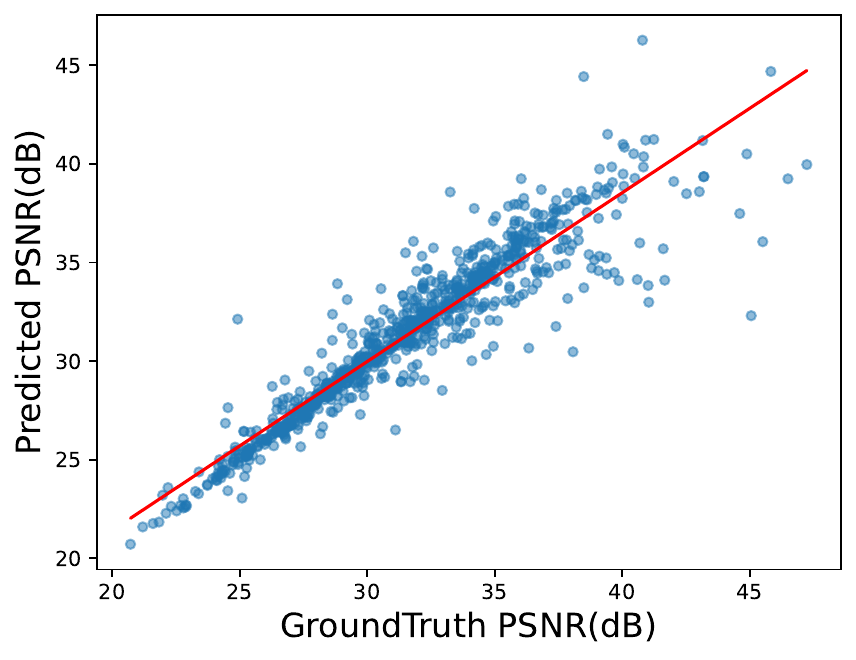}
        \label{fig:sub8.2}}
        \subfloat[DT-JRD]{
        \includegraphics[width=0.32\textwidth]{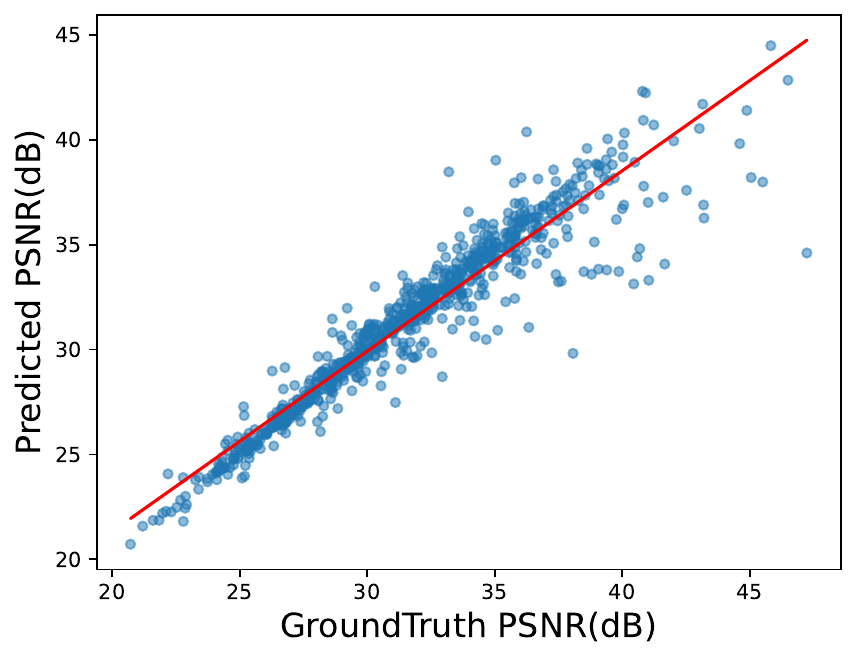}
        \label{fig:sub8.3}}
    \captionsetup{justification=justified}
    \caption{Correlation between predicted JRD and ground truth JRD in terms of PSNR.}
	\label{PSNR}
\end{figure*}

\setlength{\tabcolsep}{2.5pt} 
\begin{table}[t]
	\centering
	\caption{Ablation Studies of Key Modules in DT-JRD.}
	\begin{tabular}{ccccccccc}
		\bottomrule[1.5pt]
		\multicolumn{3}{c|}{Learning Strategies} & \multicolumn{2}{c|}{Output Features} & \multicolumn{2}{c|}{Position Embedding} & \multicolumn{1}{c}{\multirow{2}[2]{*}{$E_A$}} & \multicolumn{1}{c}{\multirow{2}[2]{*}{$E_{[27,51]}$}} \bigstrut[t]\\
		\multicolumn{1}{c}{LP} & \multicolumn{1}{c}{FF} & \multicolumn{1}{c|}{DAFT} & \multicolumn{1}{c}{Class} & \multicolumn{1}{c|}{AP}   & \multicolumn{1}{c}{Cosine} & \multicolumn{1}{c|}{Interpolation} &       &  \bigstrut[b]\\
		\hline
		\multicolumn{1}{c}{\checkmark} &       &      &       & \checkmark     &       & \multicolumn{1}{c}{\checkmark} & 6.643 & 4.849 \bigstrut[t]\\
		& \multicolumn{1}{c}{\checkmark} &       &       & \checkmark     &       & \multicolumn{1}{c}{\checkmark} & 5.728 & 4.408 \\
		&       & \checkmark     &       & \checkmark     &       & \multicolumn{1}{c}{\checkmark} & \textbf{5.574} & \textbf{4.204} \\
		&       & \checkmark     & \multicolumn{1}{c}{\checkmark} &       &       & \multicolumn{1}{c}{\checkmark} & 5.628 & 4.264 \\
  		&       & \checkmark     &  & \multicolumn{1}{c}{\checkmark}      &       & \multicolumn{1}{c}{\ding{55}} & 5.651 & 4.264 \\
		&       & \checkmark     &       & \checkmark     & \checkmark     &       &  6.032     	 &  4.458 \bigstrut[b]\\
		\toprule[1.5pt]
	\end{tabular}%
	\label{table3.4}%
\end{table}%
Furthermore, we analyzed the JRD prediction accuracy in terms of $\lvert\Delta$PSNR$\rvert$, $\lvert\Delta$SSIM$\rvert$, $\lvert\Delta$B$\rvert$ and correlation $R^2$. Lower $\lvert\Delta$PSNR$\rvert$, $\lvert\Delta$SSIM$\rvert$, $\lvert\Delta$B$\rvert$, and higher $R^2$ represent a stronger correlation with the ground truth JRD. Table \ref{quality performance} shows that DT-JRD achieved $\lvert\Delta$PSNR$\rvert$, $\lvert\Delta$SSIM$\rvert$ and $\lvert\Delta$B$\rvert$ of 0.739, 1.118$\times 10^{-2}$, and 0.140 respectively, which are the lowest among three JRD prediction models. It indicates that the quality and coding cost of the image compressed by the DT-JRD is closer to those of the ground truth JRD image. To make it more vivid, we presented a scatter plot of the ground truth PSNR and the predicted PSNR in Fig. \ref{PSNR}. The $R^2$ correlations achieved by the EL-JRD and BC-JRD are 0.908 and 0.934, respectively. The PSNR predicted by the DT-JRD is more concentrated and the $R^2$ is 0.951, which is higher than those of the other two models. It proves that the proposed DT-JRD is more accurate in predicting the JRD.

\subsection{Ablation Studies on DT-JRD}

We also conducted ablation studies on the proposed DT-JRD, where different learning strategies, output features, and position embedding schemes were analyzed. Table \ref{table3.4} shows the $E_{A}$ and $E_{[27,51]}$ when key modules of the DT-JRD were replaced or turned off. Firstly, in terms of the learning strategies, the first three rows in Table \ref{table3.4} show that LP performs poorly due to its limited learnable parameters, and FF yields over-fitting to some extent. The proposed DAFT strategy freezes the embedding module and layer norm module thus effectively protecting the common features and enhancing the generalization ability of DT-JRD. Secondly, in terms of output features, ``Class'' and ``AP'' refer to doing classification with the class token or with average pooled patch tokens, respectively. The third and fourth rows in Table \ref{table3.4} show that the average pooling of patch tokens is more conducive to capturing features with a high correlation to the JRD attribute.
Thirdly,
The last two rows in the table compare different processing methods on position embedding. The tick in the ``Interpolation'' column indicates that interpolation will be performed on the pre-trained position embedding weights followed by fine-tuning at 384$\times$384 resolution, while the cross indicates fine-tuning at 224$\times$224 resolution without interpolation. Compared to the third row where $E_A$ is 5.574, the $E_A$ in the fifth row is 5.651. A better performance is achieved by using the position embedding interpolation at a higher resolution, such that an object image can be sliced into more available patches while maintaining position information. Finally, we compare with fine-tuning at 384$\times$384 resolution with position embedding using sine and cosine functions, denoted as ``Cosine''. As shown in the last row, the $E_A$ is 6.032 which is inferior to that using interpolation.
Despite being fine-tuned at higher resolutions, this position embedding is not able to inherit the advantages of pre-trained weights. Overall, the DAFT, AP, and interpolation are the best choices for the DT-JRD model.

\subsection{Coding Performance of DT-JRD based VCM}
\begin{figure}[t]
	\centering
    \captionsetup{justification=justified}
	\includegraphics[width=0.45\textwidth]{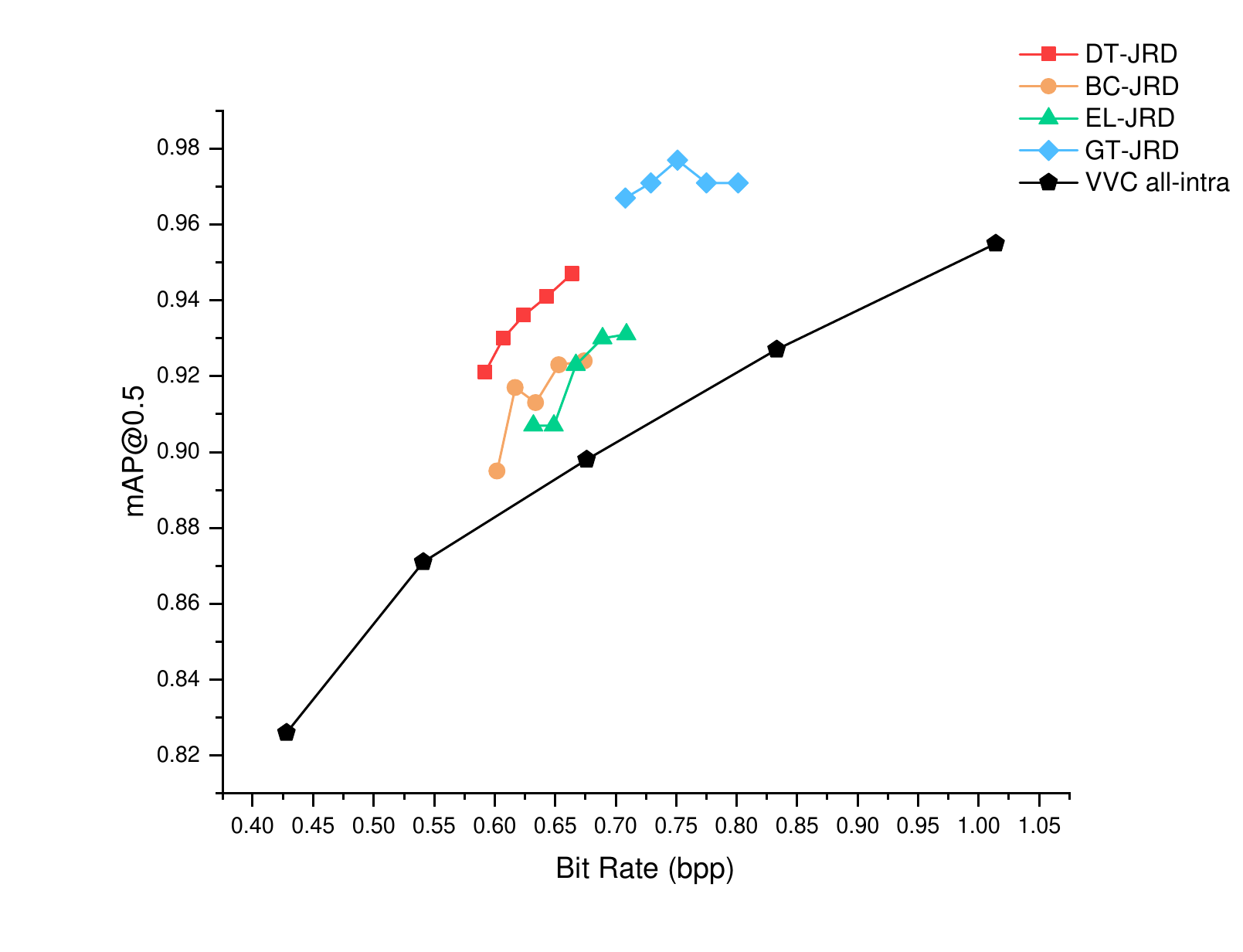}
	\caption{Relationship between mAP@0.5 and average bit rate for different coding methods.}
	\label{figmap0.5}
\end{figure}
\begin{figure}[t]
	\centering
    \captionsetup{justification=justified}
	\includegraphics[width=0.45\textwidth]{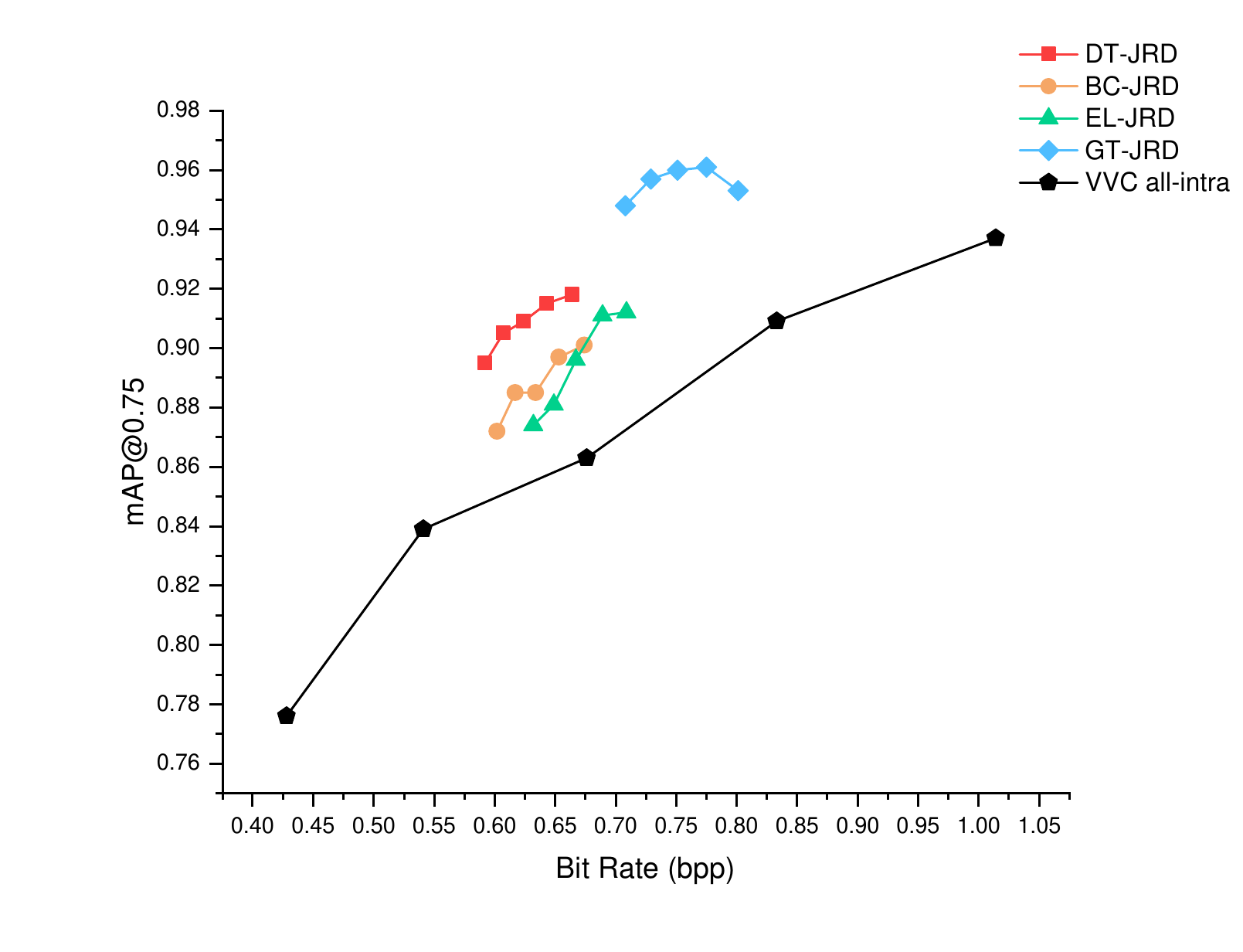}
	\caption{Relationship between mAP@0.75 and average bit rate for different coding methods.}
	\label{figmap0.75}
\end{figure}

\setlength{\tabcolsep}{4pt}
\begin{table}[t]
  \centering
  \caption{Performance comparison in terms of $R^2$ and differences of PSNR[dB], SSIM, bit rate [bpp]}
    \begin{tabular}{ccccc}
    \bottomrule[1.5pt]
    Model & $\lvert\Delta$PSNR$\rvert$ $\downarrow$ &
    $\lvert\Delta$SSIM$\rvert$ $\downarrow$ & $\lvert\Delta$B$\rvert$ $\downarrow$ &
    \multicolumn{1}{c}{$R^2$ of PSNR}$\uparrow$\\
    \hline
    EL-JRD\cite{zhang2021just} & 1.072 & 1.649$\times10^{-2}$ & 0.166 & 0.908\\
    BC-JRD\cite{zhang2023learning} & 0.890 & 1.352$\times10^{-2}$ & 0.152 & 0.934 \\
    DT-JRD  & \textbf{0.739} & \textbf{1.118$\times10^{-2}$} & \textbf{0.140} & \textbf{0.951} \\
    \toprule[1.5pt]
    \end{tabular}
  \label{quality performance}
\end{table}%

In addition to the JRD prediction accuracy, we also evaluated the coding gains achieved by the proposed DT-JRD based VCM. Fig. \ref{figmap0.5} and Fig. \ref{figmap0.75} illustrate the rate-accuracy comparison between our method and other benchmark methods, where the horizontal axis is bit rate in terms of bits per pixel (bpp) and the vertical axis represents ``mAP@0.5" and ``mAP@0.75", respectively. 
First of all, we observe that the rate-accuracy curve for VVC all-intra is at the bottom, and the mAP increases as the coding bit rate increases.
The proposed DT-JRD based VCM achieves Bjontegaard Delta Bit Rate (BDBR) savings of 29.58\% and 26.84\% with respect to mAP@0.5 and mAP@0.75, respectively. In terms of Bjontegaard Delta (BD)-mAP@0.5 and BD-mAP@0.75, the proposed DT-JRD based VCM accomplishes gains of 4.57\% and 4.99\%, which are significant over the VVC all-intra.
Furthermore, compared to the EL-JRD and BC-JRD based VCM, the proposed DT-JRD based VCM method achieves better rate-accuracy performance in both mAP@0.5 and mAP@0.75 metrics.
Specifically, Compared with the BC-JRD based VCM, the proposed DT-JRD based VCM achieves an average of 8.48\% and 9.50\% bit rate reduction concerning mAP@0.5 and mAP@0.75, respectively. In terms of BD-mAP@0.5 and BD-mAP@0.75, the proposed DT-JRD based VCM realizes gains of 2.27\% and 2.41\%. 
Similarly, compared to EL-JRD based VCM, the proposed DT-JRD based VCM achieves an average of 10.60\% and 10.36\% bitrate savings in terms of BD-mAP@0.5 and BD-mAP@0.75, respectively.
Finally, the predicted JRD by the DT-JRD is averagely larger than the ground truth JRD for about 1.67. Therefore, the rate-accuracy curve of the DT-JRD based VCM is located in the lower left part of the VCM using the GT-JRD, which indicates there is still room to be improved. 

\begin{figure}[t]
	\centering
    \captionsetup{justification=justified}
	\includegraphics[width=0.48\textwidth]{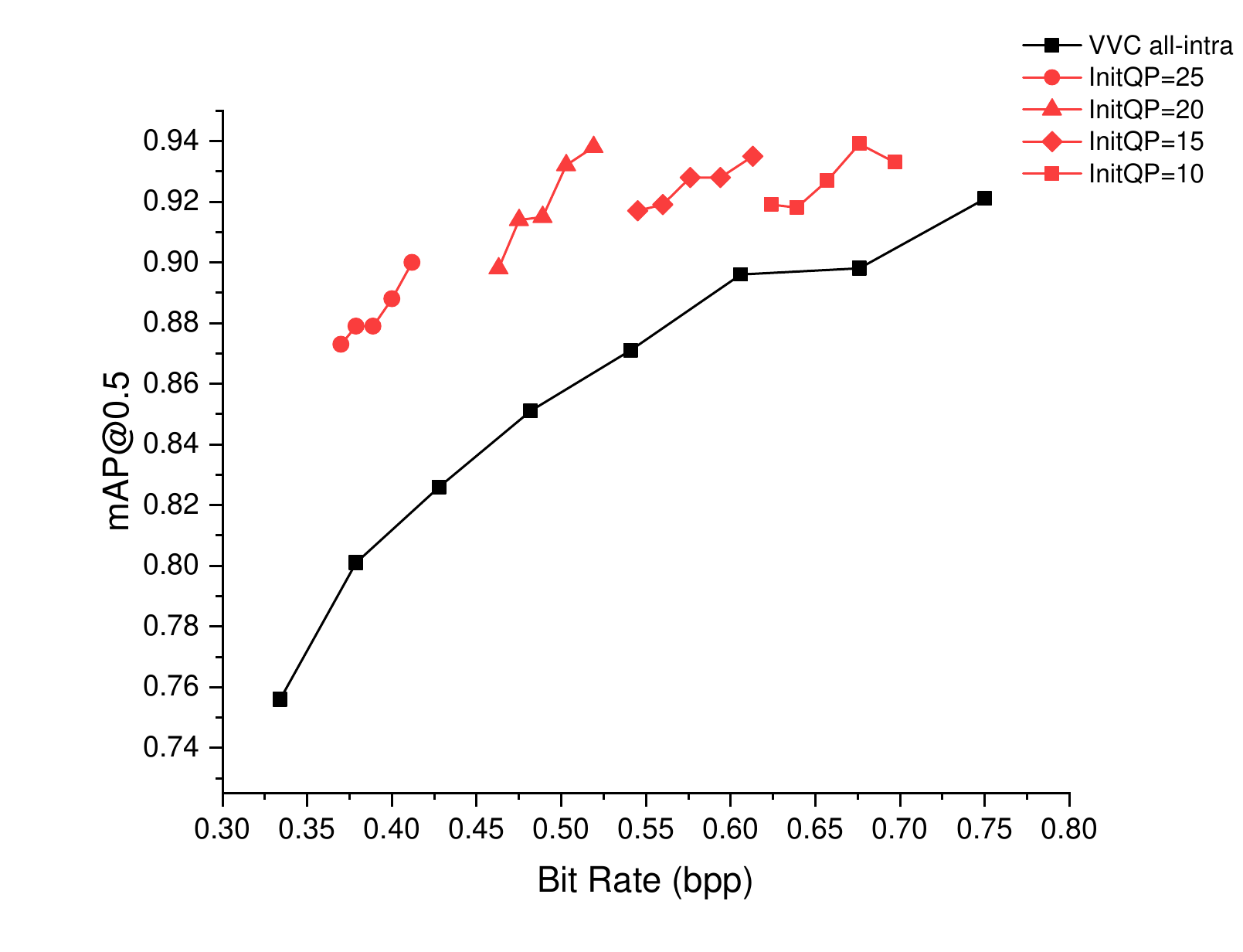}
	\caption{The relationship between mAP@0.5 and bit rate for the proposed DT-JRD based VCM using different initial QPs.}
	\label{RA}
\end{figure}

Fig. \ref{RA} illustrates the rate-accuracy curves of the proposed DT-JRD based VCM over a wider range of bit rates, where mAP@0.5 is selected as the evaluation metric. The QP of the original VVC all-intra is from 28 to 35. The proposed DT-JRD picks four sets of initial QPs $\in \{10, 15, 20, 25\}$, each of which has five offset points. When the offset is 0 and the initial QP is set to 10, 15, 20, and 25, the corresponding mAP@0.5 values of the DT-JRD based VCM are 87.3\%, 89.8\%, 91.7\%, and 91.9\%, respectively. From this perspective, compared to VVC all-intra, the DT-JRD based VCM achieves an average of 24.21\% bit rate saving and a 5.21\% BD-mAP@0.5 gain. The results demonstrate that the proposed DT-JRD based VCM outperforms the VVC all-intra in a wide range of bit rates.

\begin{figure*}[t]
\captionsetup{justification=justified}
\centering
        \subfloat[]{
        \includegraphics[width=0.15\textwidth]{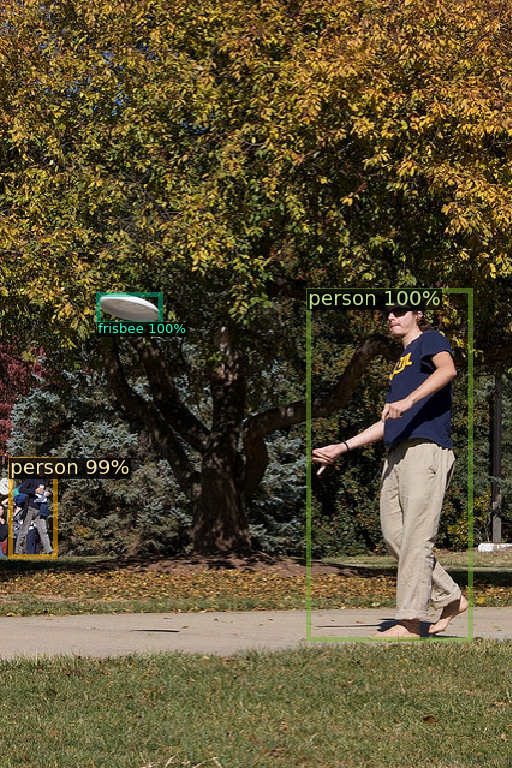}
        \label{fig:sub1.1}}
        \subfloat[]{
        \includegraphics[width=0.15\textwidth]{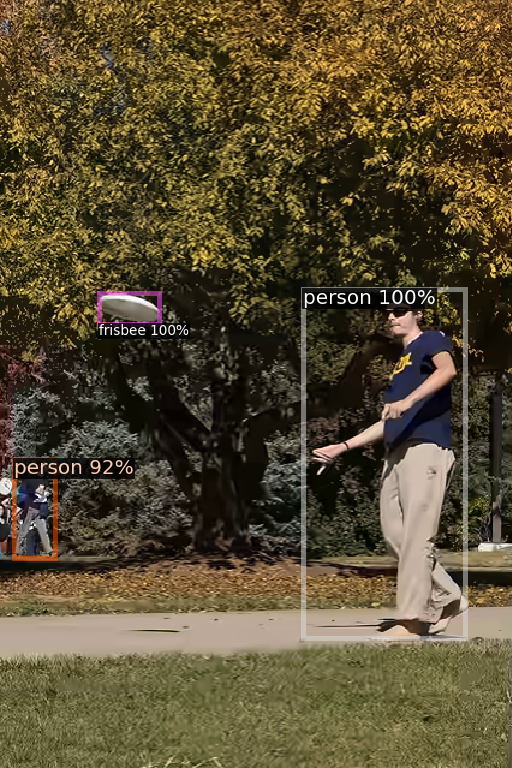}
        \label{fig:sub1.2}}
        \subfloat[]{
        \includegraphics[width=0.15\textwidth]{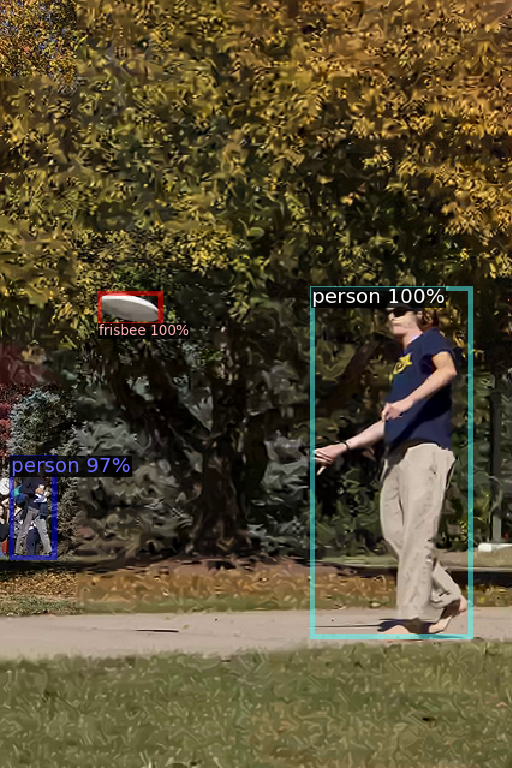}
        \label{fig:sub1.3}}
        \subfloat[]{
        \includegraphics[width=0.15\textwidth]{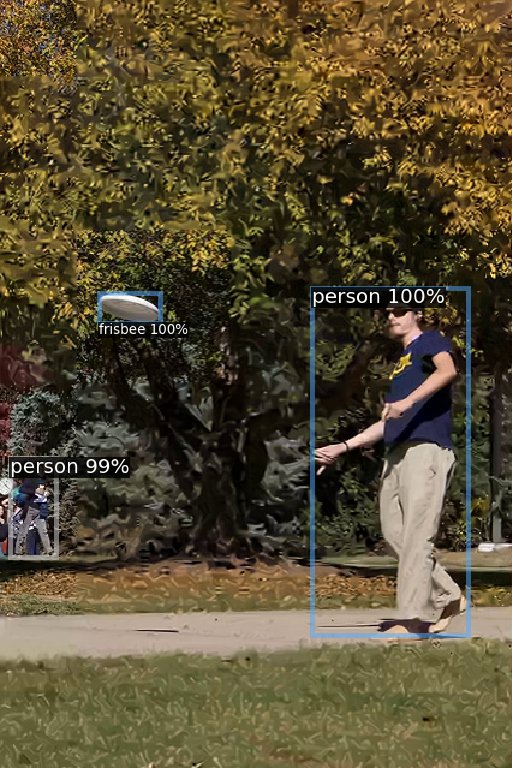}
        \label{fig:sub1.4}}
        \subfloat[]{
        \includegraphics[width=0.15\textwidth]{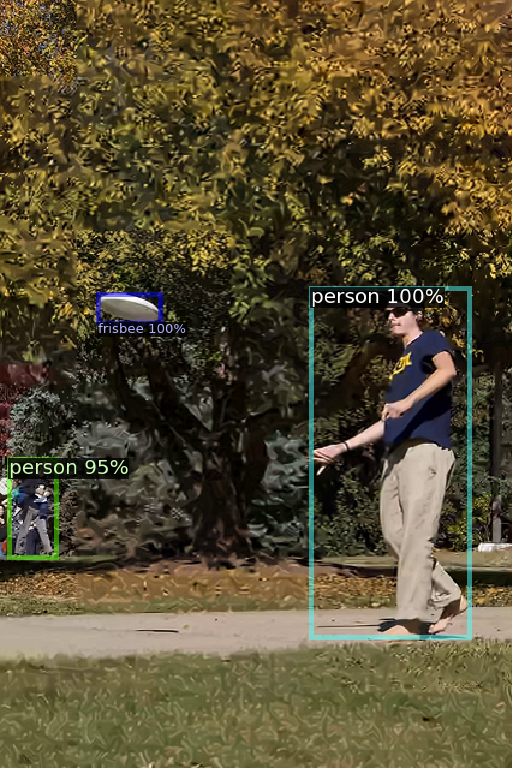}
        \label{fig:sub1.5}}
        \subfloat[]{
        \includegraphics[width=0.15\textwidth]{show_figures/00331/00331_Ours.png}
        \label{fig:sub1.6}}
        \\
        \subfloat[]{
        \includegraphics[width=0.15\textwidth]{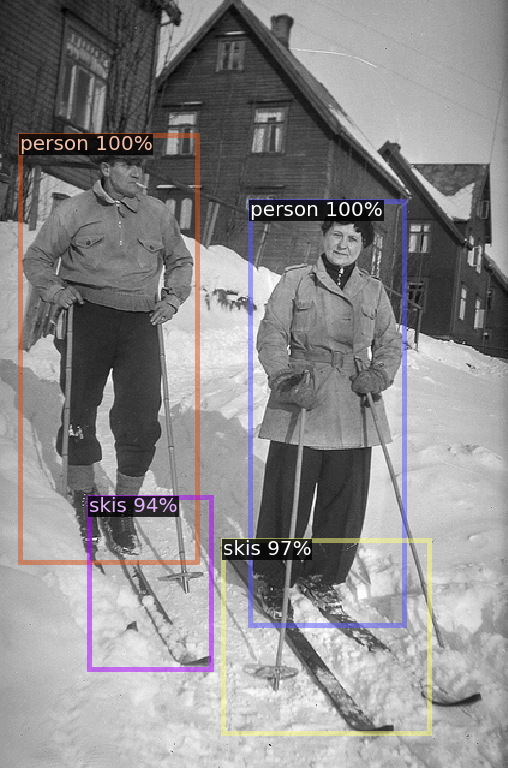}
        \label{fig:sub2.1}}
        \subfloat[]{
        \includegraphics[width=0.15\textwidth]{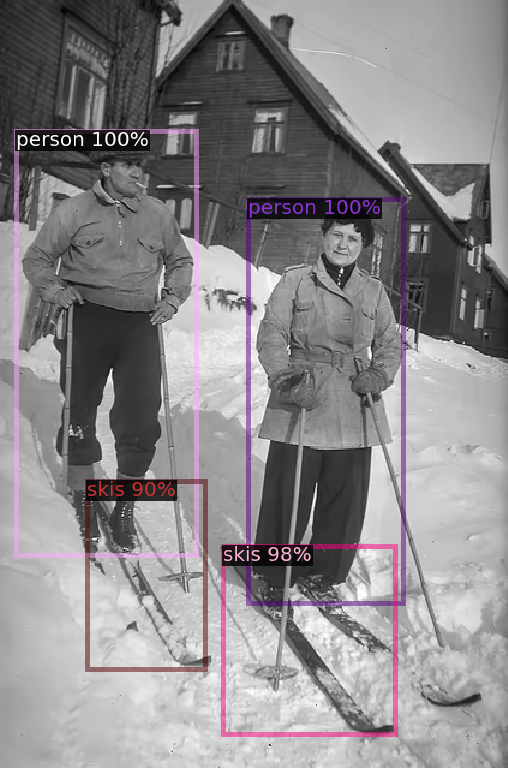}
        \label{fig:sub2.2}}
        \subfloat[]{
        \includegraphics[width=0.15\textwidth]{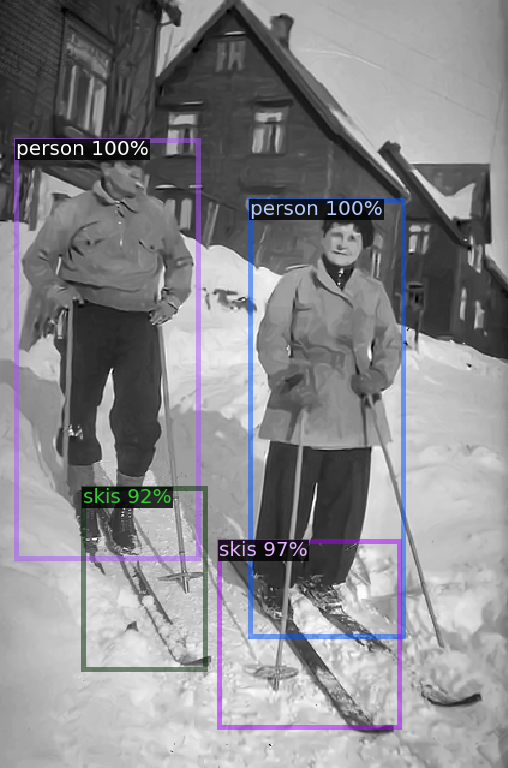}
        \label{fig:sub2.3}}
        \subfloat[]{
        \includegraphics[width=0.15\textwidth]{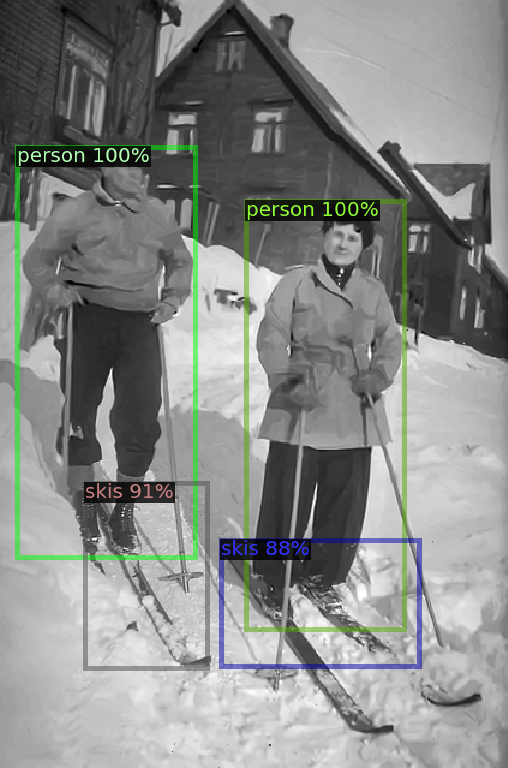}
        \label{fig:sub2.4}}
        \subfloat[]{
        \includegraphics[width=0.15\textwidth]{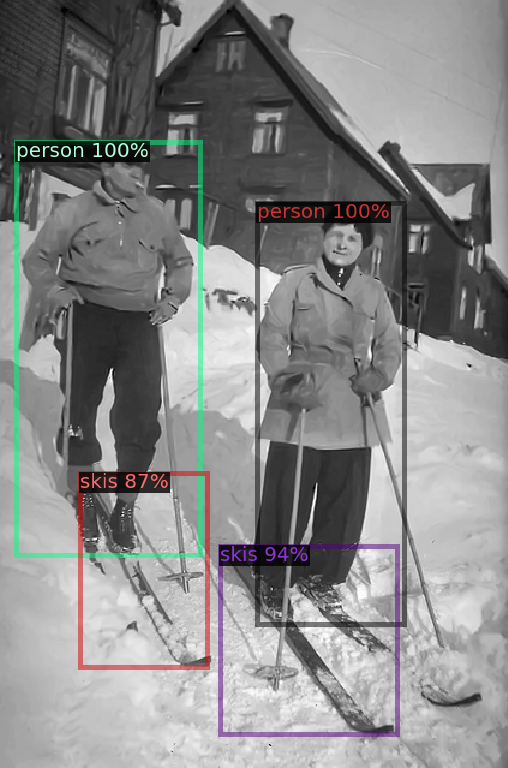}
        \label{fig:sub2.5}}
        \subfloat[]{
        \includegraphics[width=0.15\textwidth]{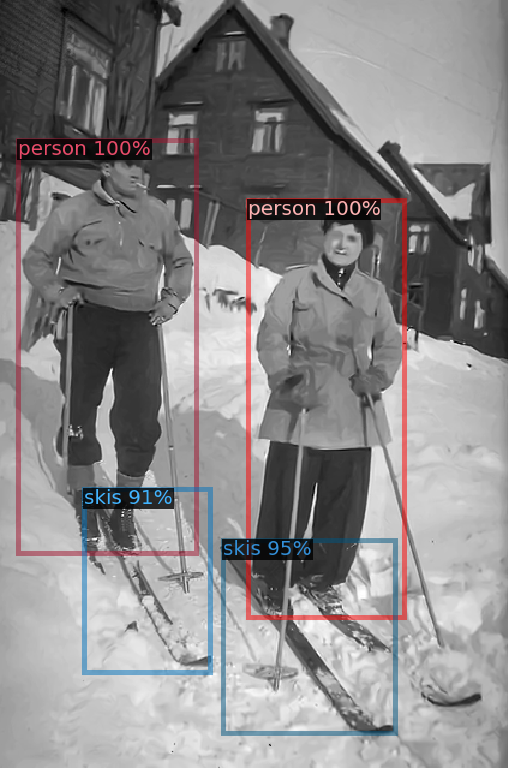}
        \label{fig:sub2.6}}
        \\
        \subfloat[]{
        \includegraphics[width=0.15\textwidth]{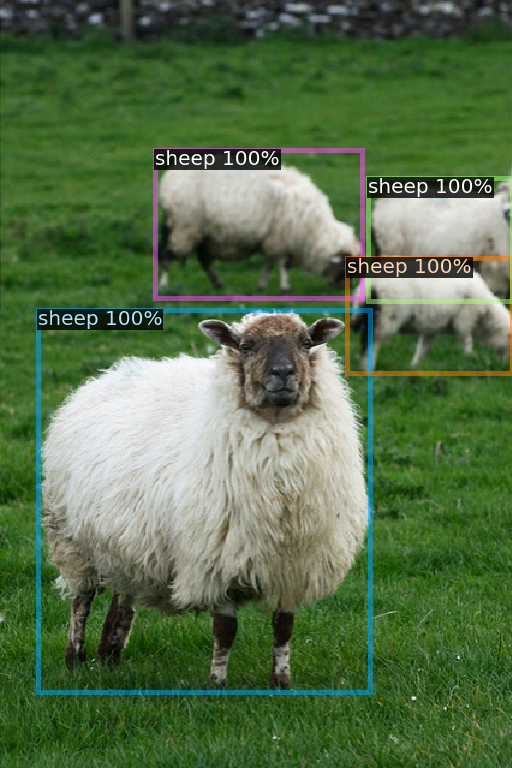}
        \label{fig:sub3.1}}
        \subfloat[]{
        \includegraphics[width=0.15\textwidth]{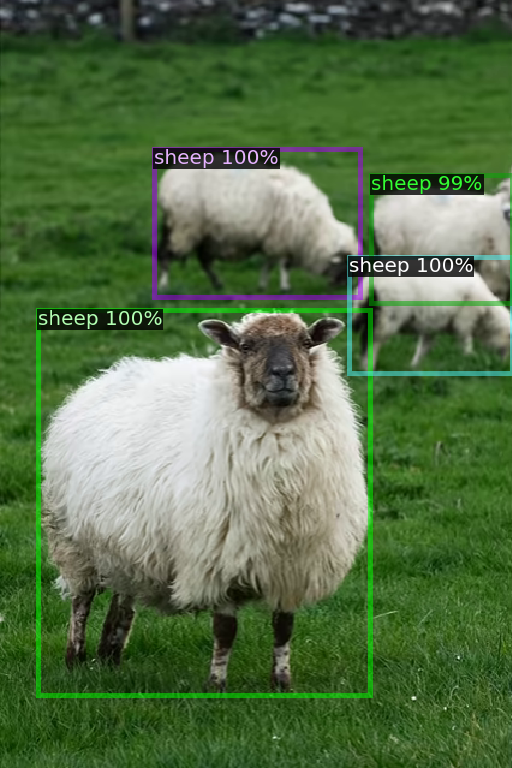}
        \label{fig:sub3.2}}
        \subfloat[]{
        \includegraphics[width=0.15\textwidth]{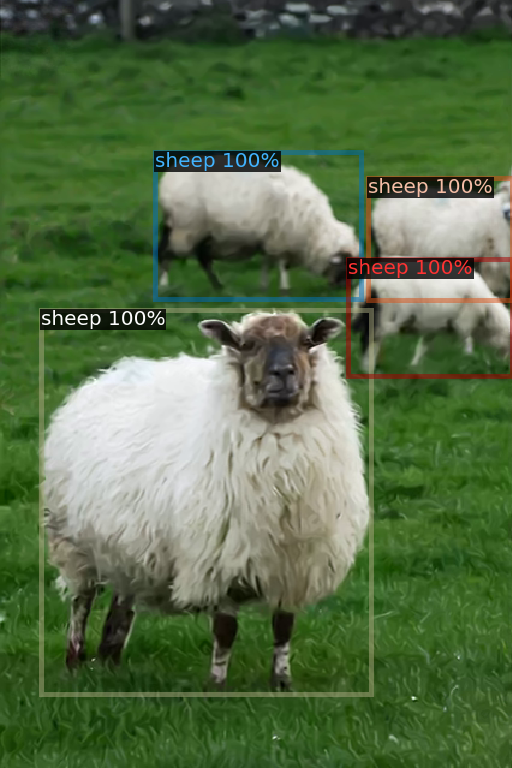}
        \label{fig:sub3.3}}
        \subfloat[]{
        \includegraphics[width=0.15\textwidth]{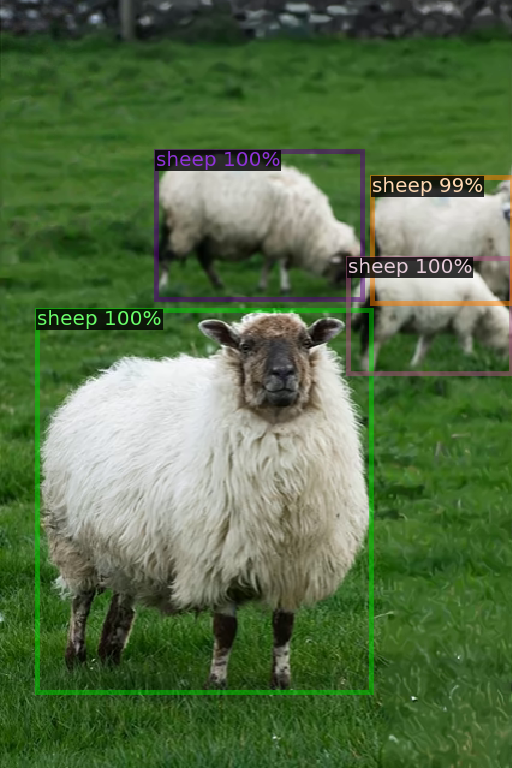}
        \label{fig:sub3.4}}
        \subfloat[]{
        \includegraphics[width=0.15\textwidth]{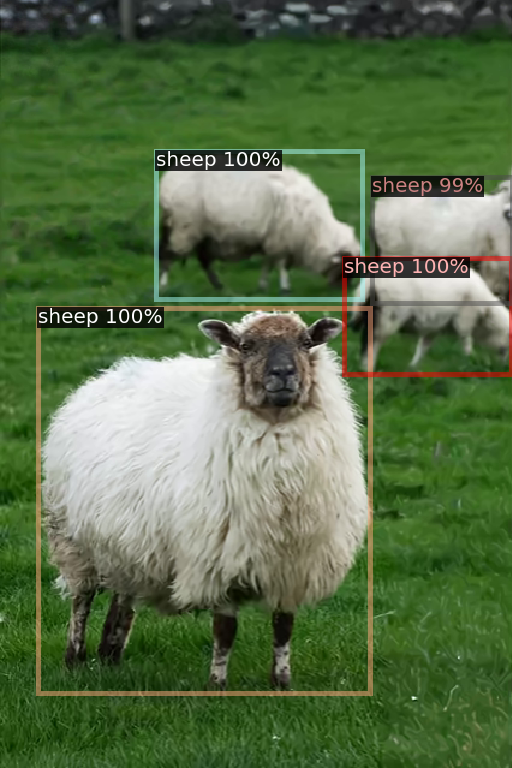}
        \label{fig:sub3.5}}
        \subfloat[]{
        \includegraphics[width=0.15\textwidth]{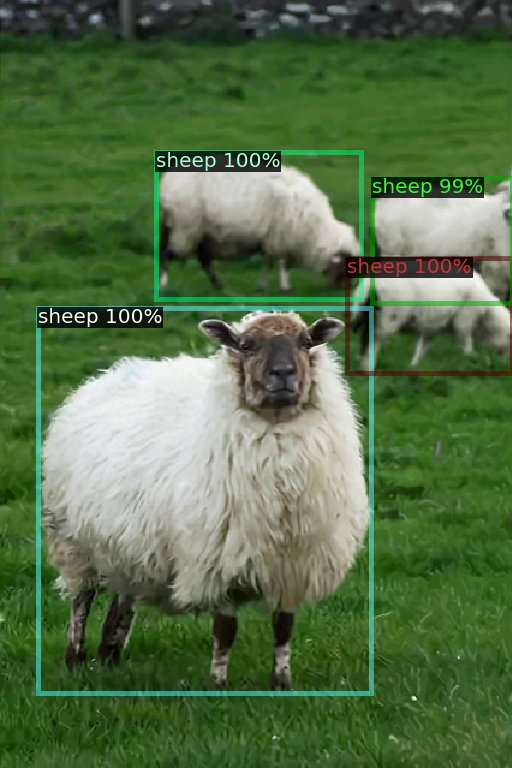}
        \label{fig:sub3.6}}
	\caption{Visualized object detection results of compressed images from different coding methods. (a)(g)(m) original images, (b)(h)(n) VVC all-intra, (c)(i)(o) GT-JRD based VCM, (d)(j)(p) EL-JRD based VCM, (e)(k)(q) BC-JRD based VCM, (f)(l)(r) DT-JRD based VCM. The bit rate of each compressed image, the average confidence, and the average IOU of the detected objects are listed as triples.
 (b) \{0.8710, 0.9725, 0.9125\}, (c) \{0.7713, 0.9888, 0.9307\}, (d) \{0.8879, 0.9942, 0.9454\}, (e) \{0.8365, 0.9810, 0.9583\}, (f) \{0.7316, 0.9831, 0.9246\}, 
 (h) \{0.7474, 0.9686, 0.8869\}, (i) \{0.5694, 0.9726, 0.9005\},
 (j)  \{0.5401, 0.9479, 0.8484\}, (k) \{0.6341, 0.9523, 0.8648\}, (l) \{0.5073, 0.9642, 0.9101\}, 
 (n) \{0.8135, 0.9972, 0.9731\}, (o) \{0.6222, 0.9976, 0.9773\}, (p) \{1.0749, 0.9975, 0.9834\}, (q) \{0.7101, 0.9959, 0.9754\}, (r) \{0.5747, 0.9955, 0.9774\}}
	\label{subfig}
\end{figure*}

\subsection{Visual Comparison for Coded Images}
We also compared the coded images from VCM using the DT-JRD, BC-JRD, and EL-JRD respectively, as shown in Fig. \ref{subfig}. The first image in each row shows the detection results of the original image. Then, from left to right, the images display the detection results of those compressed by VVC all-intra, VCM using GT-JRD, EL-JRD, BC-JRD, and the proposed DT-JRD. The coding bit rate of the compressed image, the average confidence, and the average IOU of detected objects are recorded. In the first row of images, the VCM using DT-JRD achieved a bit rate of 0.7316 bpp, which is lower than that of VVC all-intra, while obtaining an average confidence of 0.9831 and an average IoU of 0.9246, both higher than those of VVC all-intra. Additionally, the VCM using the DT-JRD outperforms the VCMs using the EL-JRD and the BC-JRD by reducing the bit rate while achieving comparable or better machine vision performance. Finally, the performance of the VCM using DT-JRD is close to that of the VCM using GT-JRD. Similar results can be observed in the images of the second and third rows. Moreover, it can be noted that the details of the foreground humans and sheep in the images exhibit distortions in human perception. However, the machine vision recognition performance remains at a high level, which indicates the DT-JRD based VCM is effective. In summary, the DT-JRD based VCM is able to significantly reduces the bit rate of transmission and storage while maintaining the accuracy of machine vision tasks.

\section{conclusions}
\label{section:conclusion}
In this paper, we propose a deep transformer based Just Recognizable Difference (JRD) prediction model for Video Coding for Machines (VCM), called DT-JRD, which reduces coding bit rate while maintaining machine vision task accuracy.
Firstly, we model the JRD prediction as multi-class classification and propose the DT-JRD framework, including
embedding module, content and distortion feature extraction and multi-class classification.
Secondly, we propose an asymptotic JRD loss by using Gaussian Distribution-based Soft Labels (GDSL), which extends the number of training labels and relaxes classification boundaries.
Finally, we propose an efficient DT-JRD based VCM where the predicted JRD is used to guide the bit allocation.
Extensive experiments demonstrate the superiority of the proposed DT-JRD model and bit reduction of the DT-JRD based VCM, which can promote machine-to-machine visual communication.
\bibliographystyle{IEEEtran}
\bibliography{IEEEabrv,reference}

\vfill
\end{document}